\documentclass[10pt, conference, letterpaper]{IEEEtran}
\IEEEoverridecommandlockouts
\usepackage{cite}
\usepackage{amsmath,amssymb,amsfonts}
\usepackage{mathtools}
\usepackage{algorithmic}
\usepackage{graphicx}
\usepackage{textcomp}
\usepackage{xcolor}
\usepackage{subcaption}
\def\BibTeX{{\rm B\kern-.05em{\sc i\kern-.025em b}\kern-.08em
    T\kern-.1667em\lower.7ex\hbox{E}\kern-.125emX}}

\newtheorem{theorem}{Theorem}[section]

\newtheorem{lemma}[theorem]{Lemma}
\newtheorem{proposition}{Proposition}[section]
\newtheorem{definition}{Definition}

\begin{document}

\title{Opportunistic Routing in Quantum Networks\\
}


\author{\IEEEauthorblockN{Ali Farahbakhsh and Chen Feng}
\IEEEauthorblockA{School of Engineering\\University of British Columbia (Okanagan Campus)\\Kelowna, British Columbia, Canada, V1V 1V7\\
Emails: ali.farahbakhsh@ubc.ca, chen.feng@ubc.ca}
}

\maketitle

\begin{abstract}
Unlike classical routing algorithms, quantum routing algorithms make use of entangled states---a type of resources that have a limited lifetime and need to be regenerated after consumption.
In a nutshell, quantum routing algorithms have to use these resources efficiently, while optimizing some objectives such as the total waiting time.
Current routing algorithms tend to keep a routing request waiting until all of the resources on its path are available.
In this paper, we introduce a new way of managing entanglement resources in an opportunistic fashion: 
a request can move forward along its path as soon as possible (even if some resources on its path are not ready).
We show that this opportunistic approach is fundamentally better than conventional approaches. In particular, our results indicate that this new approach achieves a 30-50\% improvement in the average total waiting time and average link waiting time
compared with several  state-of-the-art routing algorithms.
As a by-product of this work, we develop a new simulator for quantum routing, which can be used to evaluate various design choices under different scenarios.

\end{abstract}


\section{Introduction}
Quantum systems are able to offer superior advantages 
for certain applications (e.g., \cite{QKD, shor, QuantumSol,FastQuantByzAgr,QuanMultiValByzAgreeDDimenEntang, QuantByzAgreeSingleQutrit}).
These applications either do not have a classical solution, or their classical solutions have significantly worse performance.
The spectacular laws of quantum mechanics allow quantum systems to be inherently more suitable for certain complex problems.

\par
An indispensable part of most of these quantum systems is a quantum network.
Quantum networks are used to transfer quantum information between quantum computers, and
this information is conveyed through qubits, which are the quantum counterparts of bits.
The promising performance and applications of quantum networks serve as the building ground for a future quantum Internet~\cite{quantumInternet}.

\par
Unfortunately, the same laws of quantum mechanics  also create spectacular challenges specific to quantum systems (see, e.g.,~\cite{NoCloning}).
These challenges make quantum networks fundamentally different than their classical counterparts, and require new methods to deal with them.
Technological limitations, noise, and interaction with the environment are also among the main reasons for these challenges~\cite{DistRoutQuantumInternet,FigureMerit, ToolsQuantumDesign, opticalComm}.
Specifically, the state of a quantum network is constantly changing.
This is partly due to the fact that the links of a quantum network have finite lifetimes, and partly because their generation is probabilistic.

\par
The highly dynamic nature of quantum networks makes routing in such networks highly non-trivial.
Traditional (classical) routing algorithms rely on the robustness of the network to a great extent, and then design recovery procedures in which changes in the state of the links and the nodes are effectively handled.
In quantum networks, the state is constantly changing, so we need routing algorithms that deal with this inherent dynamism as the main issue rather than a side case.
The basis of such algorithms would be an efficient approach towards allocating the limited and temporary resources of a quantum network to the incoming requests, while managing the constant regeneration of the resources.

\par
Most of the existing quantum routing algorithms, such as~\cite{DistRoutQuantumInternet,ConcurrentRouting,RoutingEntanglementInternet}, make a design choice that adversely affects the overall performance.
Basically, existing routing algorithms wait for all of the resources (i.e., links) a request needs to be ready and then start to forward the request on its selected path.
This design choice is conservative and leads to inefficient usage of the network resources.
We argue that a more opportunistic approach better suits the unique characteristics of quantum networks.
In particular, we believe that a request should be forwarded along the path as soon as it is possible to move forward, even if it is a single hop.
As we will show in this paper, such opportunism increases the efficiency of any routing algorithm by decreasing the average total time required for a group of requests to reach their destinations.

\par
To the best of our knowledge, we are the first to demonstrate and analyze opportunism in quantum networks.
In particular, we have implemented three existing routing algorithms and demonstrated the fundamental superiority of opportunism by using both theoretical analysis and simulations.
Moreover, we have shown that opportunism can be added to almost any routing algorithm as an add-on.
Finally, we have designed and implemented a simulator that can serve as a common ground for implementing several routing algorithms with various design choices.
In fact, all three of the mentioned algorithms have been implemented with this simulator.
We will open-source our simulator, making it a useful tool for quantum routing research.
Our simulator has integrated the essential design choices of quantum routing, such as multipath routing~\cite{RoutingEntanglementInternet, EffectiveRoutingDesign}, having recovery paths~\cite{ConcurrentRouting}, and having contention-free paths.

\par
In summary, our contributions are as follows:
\begin{itemize}
    \item We introduce and analyze opportunism as a new perspective for managing quantum networks, and demonstrate its superiority with both theoretical analysis and simulations.
    \item We provide a simulator that can be used to implement different quantum routing protocols since it provides common functionalities and many design choices.
    It can also be used to compare state-of-the-art protocols through different metrics and setups.
\end{itemize}

\par
The rest of the paper is organized as follows. Section~\ref{sec: quantum networks} provides an overview of quantum networks and quantum routing. Section~\ref{Sec: opportunistic routing} delves into opportunistic routing. Section~\ref{sec: system model} describes our system model. Section~\ref{Sec: analysis} presents our theoretical analysis. The simulation results, together with a brief explanation of our simulator, are presented in Section~\ref{Sec: simulations}. Finally, the related work, and conclusion and future work are presented in Sections~\ref{Sec: related work}, and~\ref{sec: conclusion}, respectively.

\section{Quantum Networks}\label{sec: quantum networks}
In this section, we review some basics of quantum networks.

\subsection{Components}

Similar to many other networks, a quantum network consists of \emph{nodes} and \emph{links}. The nodes of a quantum network are quantum computers, which are connected through both classical and quantum channels (such as fiber optics).\footnote{The nodes of a quantum network may also include quantum repeaters~\cite{repeaters}, which facilitate long-distance quantum communication. For simplicity, we assume quantum computers have this functionality as well.}
The links of a quantum network are entangled qubit pairs (between two nodes), because they play a crucial role in quantum communication as explained below.

The quality of qubit transmission over physical links degrades exponentially as the length of the link increases~\cite{FundamentalRepeaterless,achievable,FigureMerit,ToolsQuantumDesign,rateLossTradeoff}. Therefore, \emph{quantum teleportation}
\cite{Teleport} is used to ``transfer'' quantum information.
More specifically, during quantum teleportation, an entangled qubit pair between two nodes is consumed to teleport a qubit between them.
This qubit is not sent via a physical link but is obtained at the destination through a series of measurements and classical communication.

\begin{figure}[b]
\centerline{\includegraphics[scale=1]{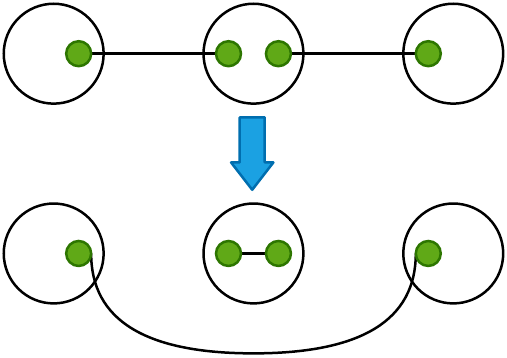}}
\caption{A simple swapping step. The big circles are quantum nodes, and the little circles are qubits. Lines represent entanglement (i.e., links).}
\label{fig: swapping}
\end{figure}

Generalizing quantum teleportation, \emph{entanglement swapping}~\cite{Swap} enables qubit transfer between two arbitrarily distant nodes by using a chain of intermediate nodes which perform a series of measurements and operations. Basically, a simple swapping protocol consumes entangled qubits between two nodes with an intermediate node to establish entanglement between these two nodes.
Fig.~\ref{fig: swapping} illustrates a simple swapping scenario.
Several protocols have been proposed for entanglement swapping (e.g., ~\cite{ToolsQuantumDesign, swap1, DistanceIndependent}).


To sum up, entangled pairs indeed play a crucial role in quantum networks, serving as the basic unit 
of resource required for successful quantum communication. This explains why they are considered as links of a quantum network. Throughout this paper, we will use the terms ``generating'' or ``establishing'' a link to refer to creating a pair of entangled qubit pairs between two adjacent nodes.


\subsection{Limitations}\label{Sec: limitations}
Quantum networks are generally hard to manage due to the following limitations.

\subsubsection{Nodes} Current prototypes for quantum computers have rather limited storage capacity for qubits~\cite{TenQubit}. This limits the number of entanglements (i.e., links) they can establish with other nodes in the network.

\subsubsection{Qubits} Qubits have different quality metrics, such as fidelity~\cite{quantumInfoTheory}.
Generally, the quality of a qubit decreases over time (due to noise and interaction with the environment). A qubit may become useless after a certain threshold,
which depends on the application (e.g., entanglement distillation~\cite{distillation}).
In other words, qubits, and therefore links, have finite lifetimes.

\subsubsection{Links} As explained before, links of a quantum network have finite lifetimes.
Moreover, the process of generating a link is probabilistic~\cite{ConcurrentRouting,DistRoutQuantumInternet,FigureMerit}.
In each attempt, there is a probability of failure.

\subsubsection{Swapping} Similar to link generation, each swapping attempt is also probabilistic with a certain failure probability~\cite{ToolsQuantumDesign}.

\subsection{Quantum Routing}\label{Sec: quantum routing}
With quantum teleportation and swapping, quantum communication can be carried out in a line network.
However, a quantum network may have a more complicated topology, making the transfer of qubits nontrivial. To cope with such complexity as well as the dynamic nature of a quantum network as explained in Section~\ref{Sec: limitations}, a typical quantum routing algorithm contains the following stages \cite{ConcurrentRouting}:

\begin{enumerate}
    \item \textbf{Reception:} First, a number of routing requests are generated in the network.
    \item \textbf{Path Selection:} Based on certain rules (see, e.g., \cite{PathSelectionQuanRepeater}), several paths are selected for every request, either offline~\cite{ConcurrentRouting} or on the fly. The links of these paths are usually reserved for the corresponding request.
    \item \textbf{Generation:} Then, links are generated along the paths for every request.
    \item \textbf{Forwarding:} Finally, once all of the required links are ready along a path, the corresponding request is forwarded based on a certain swapping pattern.
\end{enumerate}

Indeed, most existing routing algorithms follow the above order with few exceptions having a little modification. 
For example, the algorithm proposed in \cite{RoutingEntanglementInternet} 
performs the ``Path Selection" stage after the ``Generation" stage (whereas many other algorithms do the opposite).

Unlike most of the existing algorithms, our proposed approach is not related to specific details like how the paths are selected, and is instead more related to efficient management of the resources.
We achieve this by proposing a new forwarding stage for any type of routing algorithms.
Specifically, as explained in Section~\ref{Sec: opportunistic routing} and demonstrated in Section~\ref{Sec: simulations}, our proposed approach, which is opportunistic in nature, can be used as an add-on to many existing routing algorithms.





\section{Opportunistic Routing}\label{Sec: opportunistic routing}

In this section, we give a high-level overview of our opportunistic routing. Our key observation is the following: Most existing routing protocols conduct the forwarding stage rather conservatively. More specifically,
a typical routing algorithm (e.g.,~\cite{ConcurrentRouting,RoutingEntanglementInternet,DistRoutQuantumInternet}) often waits until \emph{all} of the links in a path are ready and then forwards the request by initiating the swapping process.
We believe that such a conservative approach fails to exploit the highly dynamic nature of quantum networks for the following reasons.

First, since links have finite lifetimes, it is inefficient to wait for all of the links of a particular path to get established. For example, some already  established links may time out even before the remaining links get established, leading to a waste of resources.
Second, when a request reserves the links of its path, these links cannot be reserved by other requests.
Third, since swapping is also probabilistic and it takes time to finish, by waiting for all of the links to get generated, we unnecessarily lose the opportunity to do some of the swappings while the generation is still not finished.
Therefore, by being less conservative, we can enable two types of gain.
By leveraging the resources more efficiently, we can allow every request to reach its destination faster, and at the same time prevent it from being a hindrance to other requests.

Based on the above explanations, we propose a modified forwarding stage as follows:
\begin{enumerate}
\setcounter{enumi}{3}
    \item \textbf{Forwarding:} The request is forwarded along the path as soon as it is possible, even if it is only a single hop.
\end{enumerate}

\par
\begin{figure}[htbp]
\centerline{\includegraphics[scale=0.7]{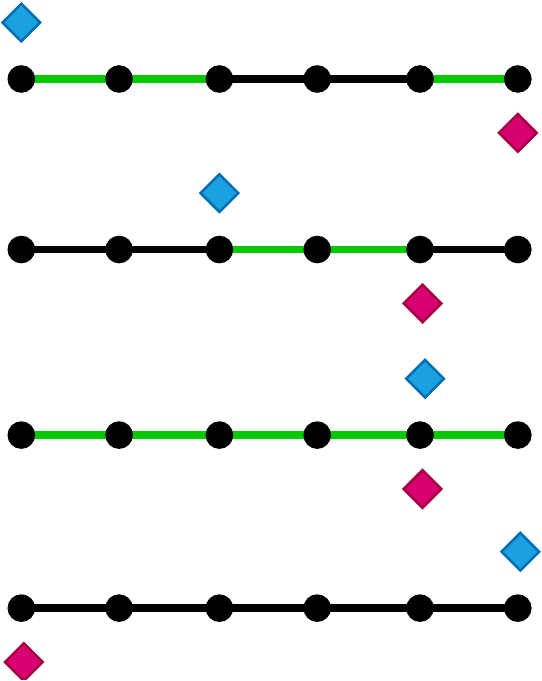}}
\caption{A simple opportunistic routing example. The green color on a link means that it is generated. Note that links support two-way qubit transfer.}
\label{fig: simple opp example}
\end{figure}

Fig.~\ref{fig: simple opp example} shows an example of being opportunistic in a simple scenario.
The blue (pink) request is going from the leftmost (rightmost) node to the rightmost (leftmost) node.
At each step, a random number of links are generated (green links), and then the requests consume the generated links, restoring them to the non-generated (black) state.
Instead of waiting for all of the links to be ready, the requests take the opportunity of moving forward when they can.
For simplicity, the intermediate steps in which links consecutively fail to get generated are omitted.

\begin{figure}[htbp]
\centerline{\includegraphics[scale=0.7]{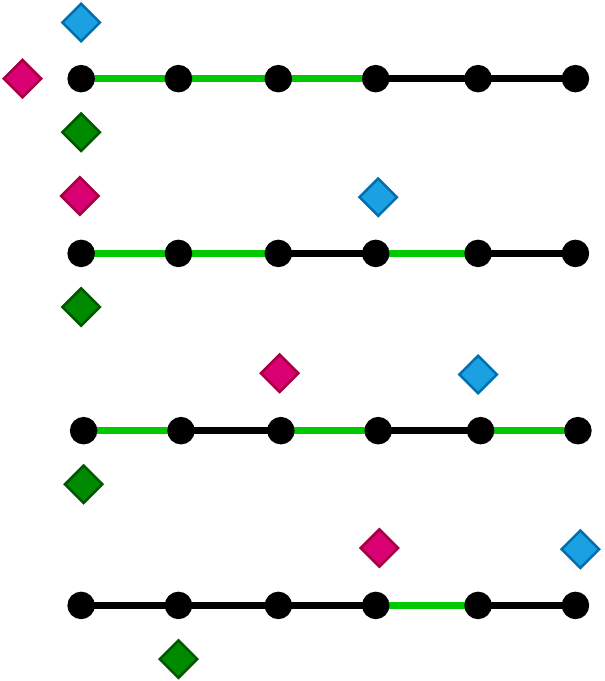}}
\caption{Another example for opportunistic routing. The swapping and reservation gains can be seen in this figure. The green color on a link means that the link is generated.}
\label{fig: N requests opp example}
\end{figure}

The gain caused by being opportunistic is worth more explanation.
To this end, imagine a scenario in which $N$ requests want to go from the same source to the same destination.
Fig.~\ref{fig: N requests opp example} shows such a scenario.
With an ordinary, non-opportunistic routing algorithm, each one of these requests has to wait for all of the links of the path to be ready, and then it gets forwarded.
This means that the requests have to wait $N$ times the time needed for a single request, on average.
However, by acting opportunistically, we can exploit the generated links as soon as it is possible for a request to move forward, and then regenerate the consumed links for the other requests.
This clearly demonstrates the two-fold gain of opportunism: 1) each request reaches the destination sooner because a part of the waiting time for the link generation is used to perform some of the swappings, and 2) resources are freed more quickly, resulting in less waiting time for requests which are waiting for the prior requests to complete transmission.
We call the first gain as the \textit{swapping} gain and the second one as the \textit{reservation} gain.

\par
There is a subtle point about the forwarding stage, which is worth mentioning.
Forwarding a request can be interpreted in two ways.
First, the request may actually be forwarded along the path, as a continuation of the swapping process.
That is, as soon as a swapping step is successful along the path, the request can be forwarded to the corresponding intermediate node using teleportation.
Alternatively, the request can wait until the swapping process is completely finished, and then it can be sent from the source to the destination using a single quantum teleportation step.
Both of these approaches can be considered opportunistic in the sense explained in the previous paragraphs, as long as the swapping process is started as soon as it is possible.
Indeed, this perspective suggests that opportunism can be leveraged in two layers: first in leveraging the generation time for a part of swapping, and then in acting opportunistically during the swapping process.
Both of these ``forwarding'' methods have their advantages and disadvantages.
With the former method, the request does not need to start from the beginning if a part of the swapping process fails, while on the other hand, it will consume a part of the corresponding intermediate node's memory, possibly harming the routing of other requests.
With the latter method, no extra resources are used for serving the request, but it has to start from the beginning every time a part of the swapping process fails.
For a neat illustration, the figures of this paper demonstrate the former method, so that requests are actually moving forward as more links are generated.
However, we have chosen the latter method for swapping in the simulations, and requests have to start from the beginning if their swapping process is not successful.

\par
As a final consideration, as explained in Section~\ref{Sec: quantum routing}, opportunism is an approach targeting the general management of quantum networks, and it can be used in any algorithm since it does not depend on specific details.
Details like how the resources are allocated and the paths are selected are not important at this level.
The point is to treat the links of a network as precious resources and use them as soon as possible.
This mindset can even be extended to be an integral part of a quantum network layer protocol (e.g.,~\cite{QuantumNetworkProtocol}).

\subsection{$k$-opportunism}
The opportunism discussed in this section is somewhat greedy and extreme: a request moves as soon as it can move.
This may not be harmful in networks with a low number of requests, but the more the requests get, the harder it gets for the network to manage the resources.
On a large scale, forwarding a request as soon as a link is available may decrease the efficiency of routing for some other request.
Additionally, requests may have different degrees of priority.
Therefore, the opportunistic approach needs to have a degree of flexibility to provide dynamic routing means for a highly dynamic quantum network.
\par
To this end, we introduce \emph{$k$-opportunism}.
Recall that the proposed forwarding stage required a request to move forward once it is possible, even for one hop.
We can change this stage as follows:
\begin{enumerate}
\setcounter{enumi}{3}
    \item \textbf{Forwarding:} The request is forwarded along the path as soon as the first $k$ immediate and consecutive hops are available.
\end{enumerate}

\begin{figure}[htbp]
\centerline{\includegraphics[scale=0.7]{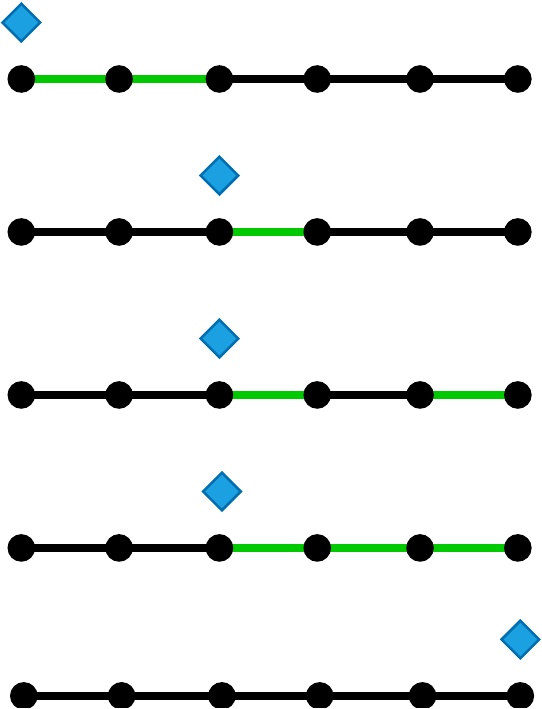}}
\caption{An example of a 2-opportunistic approach. The green color on a link means that the link is generated.}
\label{fig: 2-opportunism}
\end{figure}

Fig.~\ref{fig: 2-opportunism} shows an example of a 2-opportunistic approach, in which a request is going from the leftmost node to the rightmost node.
Here, the request does not move forward unless the two immediate and consecutive links are ready.
Intuitively, this new opportunism provides us with a spectrum of choices, allowing us to impose a distributed control over every request's greediness.
Indeed, the extreme opportunism introduced at the beginning of Section~\ref{Sec: opportunistic routing} is a 1-opportunistic approach.
Now, this spectrum of opportunism, obtained by changing $k$, which we call the degree of opportunism, provides natural means for managing requests with different priorities.
That is, we can assign lower values of $k$ for requests with higher priorities.
In this way, high-priority requests will be able to leverage the links of the network sooner than the other requests, and the higher $k$ gets for a request, the lower its chances are to use a link for its transmission.

\section{System Model}\label{sec: system model}
\subsection{Network, Nodes, and Links}\label{Sec:Network}
We model a quantum network as a graph.
The nodes of the graph are quantum computers, and the edges of the graph are the links (i.e., entangled pairs).
For simplicity, we assume that at most one link can exist between any two adjacent nodes, because such a simple setup is sufficient to demonstrate the advantage of our opportunistic routing. 
Unlike~\cite{ShortcutQuanRouting,DistRoutQuantumInternet}, we do not use virtual links, i.e., entanglement between non-adjacent nodes, as a primary resource.
The reason for this choice is also similar.
We want to demonstrate that links are precious resources and have to be consumed opportunistically, and it does not matter if they are between adjacent or non-adjacent nodes.
\par
We use $(s, d)$ to denote a request. Here, $s$ is the source and $d$ is the destination. When it is clear from the context, we may also use $r$ for a single request, and $\{r_{i}\}_{i=1}^{n}$ for a group of requests.
\par
We denote the lifetime of an entangled pair (i.e., a link) as $L$.
We assume that quantum nodes have infinite memories.

\subsection{Entanglement Generation and Swapping}
We assume that every entanglement generation attempt between adjacent nodes succeeds with probability $p_\textsc{gen} = p$.
Entanglement generation for a link is continuously attempted until the link
is generated.
Furthermore, once a link is consumed or timed out, it starts the generation process again.
Finally, we assume that every swapping attempt succeeds with probability $p_\textsc{swap}$.

\subsection{Waiting Time}
We divide the waiting time of a request into two parts as follows.
\begin{itemize}
    \item \textbf{Swapping:} It is the time required for a swapping process over a given number of links.
    \item \textbf{Generation:} It is the amount of time it takes for all of the links required for the path of the request to get generated.
\end{itemize}
We consider a discrete-time model, in which every time step is equal to the time it takes to attempt an entanglement generation between two adjacent nodes. 
Furthermore, we also assume that a single swapping step takes place at the end of each time slot.
That is, each request has a chance of one swapping step at the end of each slot.
The waiting times are explained in the following sections.
\subsubsection{Swapping waiting time}
Different algorithms have been suggested for entanglement swapping (e.g.,~\cite{swap1}).
Any algorithm can be selected for this process, and each would lead to a certain formula for the waiting time.
We denote the total time it takes to swap over $M$ links as $t_\textsc{swap}(M)$.
\subsubsection{Generation waiting time}
We denote the amount of time required for a path with $M$ links to be generated as $W(M, p)$.
We also denote the time required for link $i$ of this path to get established by $T_{i}$.
Note that $\{T_{i}\}$ are i.i.d. and they follow a geometric distribution with parameter $p$.
Note also that $W(M, p)$ is a random variable, and we denote $R(M,p) = E\{W(M, p)\}$.
It is shown in~\cite{FigureMerit,RateAnalysisHybrid} that, if $L = \infty$,
\begin{equation}
    R(M,p) = \sum_{k=1}^{M} {M\choose k} \frac{(-1)^{k+1}}{1 - (1 - p)^{k}}.
    \label{EQ:R_inf}
\end{equation}
Also,
note that $W(M,p) = \max_{i}\{T_{i}\}$.

\section{Analysis}\label{Sec: analysis}
We focus on the swapping and reservation gains in this section.
Our goal is to show that the opportunistic approach improves performance by decreasing the average total waiting time.
We also show the spectrum property of opportunism by analyzing $k$-opportunism.

To make our analysis tractable, 
we mainly focus on a line network here (which will be relaxed in the simulation section).
More specifically, we assume a number of quantum nodes in a single line, in which every node has two neighbors (except the endpoints, which have one).
We will call the endpoints of this network as $A$ and $B$ throughout the paper.
We assume that this line network has $M$ links.

In addition, we introduce several simplifying assumptions for our analysis.
First, we assume $L = \infty$.
That is, a generated link remains intact forever, until a request consumes it, after which the link has to be regenerated.
Second, following~\cite{DistRoutQuantumInternet,RateAnalysisHybrid,ToolsQuantumDesign} we assume that $p_\textsc{swap}=1$.
Finally, we have used a simple swapping algorithm in which the swapping operations are carried out on a one-by-one basis, starting from the source towards the destination.
For this simple scheme, we have $t_\textsc{swap}(M)=M-1$.

\subsection{A Single Request}\label{sec: analysis-swapping}
When sending a single request, the opportunistic approach helps to reduce the total waiting time by mixing the swapping and generation waiting times.
In the conventional approach, the request has to wait for the generation to finish, then it can begin the swapping process.
The opportunistic approach enables swapping and generation to happen simultaneously, leveraging the time the request has to wait for link generation to do a part of swapping.
This is the swapping gain mentioned in Section~\ref{Sec: opportunistic routing}.
For this part, let us show the waiting time of a single request in the non-opportunistic and opportunistic settings with $Q$ and $Q_\textsc{opp}$, respectively.
\par
We consider a single request $r = (A, B)$.
With the opportunistic approach, $r$ will be somewhere in the middle of the network\footnote{The request is actually in the source, but the swapping process has reached the middle of the network. See Section~\ref{Sec: opportunistic routing}.} once all of the links have been generated (i.e, after $W(M,p)$).
The number of links $r$ has been able to swap along the way until all of the links are generated is how opportunism shows itself, and indeed this number can be interpreted as the swapping gain caused by opportunism.
Let us denote the position of $r$ after $W(M,p)$ by $K$.
We are interested in $E\{K\}$.
\par
Let us use $W$ to show $W(M,p)$, for brevity.
Furthermore, let us define $W_{i} = \max_{j\leq i} \{T_{j}\}$.
To calculate $E\{K\}$, we have $E\{K\} = \sum_{k=1}^{M}P(K\geq k)$.
Note that $K \geq k$ implies that the first $k-1$ links were not the last links to get generated, and $W \geq k$ since every swapping attempt takes one time step.
We thus have $P(K \geq k) = P(W > W_{k-1}, W \geq k)$.
Let us define $\Tilde{W} = \max_{i > k-1}\{T_{i}\}$.
$W > W_{k-1}$ implies $\Tilde{W} > W_{k-1}$, and vice versa.
Therefore, it is easy to see that $P(W > W_{k-1}, W \geq k) = P(\Tilde{W} > W_{k-1}, \Tilde{W} \geq k)$.
We can decompose this event as follows $P(\Tilde{W} > W_{k-1}, \Tilde{W} \geq k) = \sum_{i=k}^{\infty} P(\Tilde{W} > W_{k-1}, \Tilde{W} = i)$, which can further be expressed as $P(\Tilde{W} > W_{k-1}, \Tilde{W} \geq k) = \sum_{i=k}^{\infty} P(W_{k-1} < i, \Tilde{W} = i)$.
Now, since $W_{k-1}$ and $\Tilde{W}$ are related to two disjoint sets of links, they are independent of each other, and we have $P(\Tilde{W} > W_{k-1}, \Tilde{W} \geq k) = \sum_{i=k}^{\infty} P(W_{k-1} < i)P(\Tilde{W} = i)$.
We can now evaluate each part separately.
First, we have $P(W_{k-1} < i) = P(\cap_{j=1}^{k-1}\{T_{j}<i\}) = \prod_{j=1}^{k-1}P(T_{j}<i)=P(T_{1}<i)^{k-1}$.
Here we have used the fact that the $T_{j}$ are i.i.d.
It is easy to see that $P(T_{1}<i) = 1 - (1-p)^{i-1}$, and letting $q = 1-p$, we have $P(W_{k-1} < i) = (1 - q^{i-1})^{k-1}$.
Now, we have $P(\Tilde{W} = i) = P(\Tilde{W} \leq i) - P(\Tilde{W} < i)$, which can be calculated very similar to the procedure above, and one would have $P(\Tilde{W} = i) = (1 - q^{i})^{M-k+1} - (1 - q^{i-1})^{M-k+1}$.
Combining these results leads us to $P(K \geq k) = \sum_{i=k}^{\infty} (1 - q^{i-1})^{k-1}[(1 - q^{i})^{M-k+1} - (1 - q^{i-1})^{M-k+1}]$.
Therefore, we have
\begin{equation}
    \begin{aligned}
        E\{K\}=& \sum_{k=1}^{M} \sum_{i=k}^{\infty} (1 - q^{i-1})^{k-1}\\
        &[(1 - q^{i})^{M-k+1} - (1 - q^{i-1})^{M-k+1}].
    \end{aligned}
\end{equation}
\par
Now, we have $P(K > 0) > 0$.
Furthermore, $r$ has to swap along $M-K$ links after $W(M,p)$, and $t_\textsc{swap}(M-K) < t_\textsc{swap}(M)$.
We have $Q_\textsc{opp} = R(M, p) + E\{t_\textsc{swap}(M-K)\}$.
Thus it is easy to see that $Q_\textsc{opp} < Q = R(M, p) + t_\textsc{swap}(M)$.
This proves the following theorem.
\begin{theorem}\label{Theorem:T1}
For a single request $S = \{(A, B)\}$, we have $Q_\textsc{opp} < Q$.
\end{theorem}
\subsection{A Group of Requests in the Same Direction}\label{Sec:NReq}
Here, we demonstrate the reservation gain caused by opportunism.
To do this, we will assume $N$ requests going from $A$ to $B$.
Note that to show opportunism decreases the total waiting time in this scenario, all one has to do is to leverage the swapping gain, as explained in Section~\ref{sec: analysis-swapping}, for every request.
However, as discussed in Section~\ref{Sec: opportunistic routing}, opportunism enables another gain, which we call the reservation gain, and this gain has a different nature than the swapping gain.
\par
To fully demonstrate the reservation gain, we will assume that swapping takes no time (i.e., $t_\textsc{swap}(M) = 0$).
This way, a request can perform as many swapping steps as it can in a single time step.
In this new setting, the swapping gain will vanish, and in the case of a single request, opportunism will not improve the performance.
However, when having multiple requests, we can show that opportunism decreases the total waiting time, thus demonstrating the existence of what we call the reservation gain.
\par
Let us use $T_{ij}$ (where $i\leq M$ and $j\leq N$) to show the time it takes to generate the $i$-th link for the $j$-th time (i.e., for the $j$-th request), once it has been consumed for the $(j-1)$-th request.
Each $T_{ij}$ follows a geometric distribution with parameter $p$, and $T_{ij}$ are independent.
Furthermore, changing the notation for a bit, let us show the amount of time required for $N$ requests to reach their destinations using opportunism with $W_N$.

Let us express $\{T_{ij}\}$ in a more compact matrix form:
\begin{equation}\label{EQ: D}
    D = \begin{pmatrix}
    T_{11} & T_{21} & \dots & T_{M1} \\
    T_{12} & T_{22} & \dots & T_{M2} \\
    \vdots & \vdots & \ddots & \vdots \\
    T_{1N} & T_{2N} & \dots & T_{MN} \\
    \end{pmatrix}.
\end{equation}
Note that $D_{ij} = T_{ji}$. This allows us to have the links on the rows of the matrix, making it look similar to a line network.
In particular, each row $j$ of $D$ shows the $j$-th generation times for the links of the network.

In order to analyze $W_N$, we need to introduce a few intermediate variables.
Let us define:
\begin{equation}\label{EQ: opportunism}
W_{ij} = \begin{cases}
\max_{k\leq i} T_{kj} & j=1 \\
\max_{k\leq i} \{W_{k,j-1}+T_{kj}\} & j > 1
\end{cases}
\end{equation}
This recursive structure allows us to have the following proposition.
\begin{proposition}\label{Pro: 1}
The total waiting time in the opportunistic setting is $W_N = W_{MN}$.
\end{proposition}
\begin{IEEEproof}
Let us show the $N$ requests with $\{r_i\}_{i=1}^{N}$.
First, we show that the proposition holds for $N=2$.
Link $i$ will start generation for $r_{2}$ once it has been generated and consumed by $r_{1}$.
It will be consumed once all of the first $i$ links are generated for $r_{1}$, so that $r_{1}$ can take the opportunity (remember that swapping takes no time).
Hence, link $i$ can start its generation for $r_{2}$ at time $W_{i1}=\max_{k\leq i}T_{k1}$.
Once this generation has taken place, now link $i$ takes $T_{i2}$ time steps to get generated for $r_{2}$, which means $r_{2}$ should wait a total amount of $W_{i1}+T_{i2}$ time slots for the $i$-th link to get generated.
Since there is no swapping gain, we thus conclude that the total waiting time is equal to $\max_{i\leq M}\{W_{i1}+T_{i2}\}$, which is the formula for $W_{M2}$.
This means that $W_2=W_{M2}$.
\par
We now assume that the lemma is true for $i\leq N-1$, and prove it is also true for $N$.
The proof is similar.
Request $r_{N}$ has to wait for $r_{N-1}$ to consume link $i$, and then link $i$ can start the generation process for the $N$-th time.
Request $r_{N-1}$ can consume link $i$ once it can consume all of the first $i$ links, and this takes $W_{i,N-1}$ by the induction hypothesis.
It takes $T_{iN}$ time slots for the $N$-th generation of link $i$ to succeed.
Thus, link $i$ is ready for $r_{N}$ at time $W_{i,N-1}+T_{iN}$, which means $r_{N}$ will have all of the links ready and reach the destination at time $\max_{i\leq M}\{W_{i,N-1}+T_{iN}\}$.
This is exactly the definition of $W_{MN}$, and leads us to $W_N=W_{MN}$.
\end{IEEEproof}

We also have the following lemma, which proves to be useful in the proofs.
\begin{lemma}\label{lem: 1}
$W_{ij}$ are ascending for every fixed $j$.
That is, if $k < i$, then $W_{kj}\leq W_{ij}$.
\end{lemma}
\begin{IEEEproof}
This lemma is an immediate result of the fact that for two finite sets $U\subset V$ and a function $f$ on these sets, we have $\max_{u\in U}f(u)\leq \max_{v\in V}f(v)$.
\end{IEEEproof}

We can proceed and prove that opportunism indeed decreases the total waiting time.
In the following, $W^{\textsc{up}}_N$ is the total waiting time in the non-opportunistic approach.
\begin{proposition}\label{Pro: 2}
Opportunism decreases the total waiting time.
That is, we have $W_N \leq W^{\textsc{up}}_N$.
\end{proposition}
\begin{IEEEproof}
In the non-opportunistic approach, each request has to wait for \emph{all} of the links to get established.
Therefore, the $j$-th request has to wait for $\max_{i}T_{ij}$ steps.
After each such step, all of the links are consumed, and the next request starts from scratch.
We thus conclude that:
\begin{equation}\label{eq: non-opportunism}
W^{\textsc{up}}_N=\sum_{j\leq N}\max_{i\leq M}T_{ij}.
\end{equation}
We now use induction on the number of requests to prove the theorem.
if $N=1$, then the two waiting times coincide, since a single request has to wait for all of the links, no matter what.
Now, assume that the claim holds for $N-1$.
The induction hypothesis then tells us that $W_{N-1}\leq W^{\textsc{up}}_{N-1}$.
We have:
\begin{align*}
W_N &= \max_{i\leq N} \{W_{i,N-1} + T_{iN}\}\leq\max_{i\leq N}W_{i,N-1} + \max_{i\leq N}T_{iN} \\
&\overset{(i)}{=} W_{M,N-1} + \max_{i\leq N}T_{iN} \overset{(ii)}{\leq}W^{\textsc{up}}_{N-1}+\max_{i\leq N}T_{iN} \overset{(iii)}{=} W^{\textsc{up}}_N
\end{align*}
in which equality (i) comes from Lemma~\ref{lem: 1}, inequality (ii) is based on the induction hypothesis and Proposition~\ref{Pro: 1}, and equality (iii) is just Equation~\eqref{eq: non-opportunism}.
\end{IEEEproof}

As explained in Section~\ref{Sec: opportunistic routing}, it turns out that opportunism has a spectrum of degrees, leading all the way to non-opportunism.
Put simply, instead of forwarding a request as soon as an immediate link is available, we can forward the request once there are $k$ immediate and consecutive links available.
In order to analyze $k$-opportunistic settings in a coherent way, we define the following variables:
\begin{equation}\label{EQ: k-opportunism}
    W_{ij}^{k} = \begin{cases}
    \max_{m\leq \min\{i+k-1, M\}} T_{mj} & j=1 \\
    \max_{m\leq \min\{i+k-1, M\}} \{W_{m,j-1}^k+T_{mj}\} & j > 1
    \end{cases}
\end{equation}
Note that Equation~\eqref{EQ: opportunism} is a special case of Equation~\eqref{EQ: k-opportunism} for $k=1$.
Based on this definition, we also denote the total waiting time in the $k$-opportunistic setting by $W_N^{k}$.
The following proposition is a straightforward generalization of Proposition~\ref{Pro: 1}.
\begin{proposition}\label{Pro: 3}
The total waiting time in the $k$-opportunistic setting is $W_N^{k}=W_{MN}^k$.
\end{proposition}

Similarly, Lemma~\ref{lem: 1} can also be extended.
The proof is exactly the same, and is omitted for brevity.
\begin{lemma}\label{lem: w^k ascending}
$W_{ij}^k$ are ascending for fixed $j$ and $k$.
That is, if $m < i$, then $W_{mj}^k\leq W_{ij}^k$.
\end{lemma}

The important point about opportunism is the fact that the waiting times form an ascending order based on the opportunism degree.
This is captured by the next proposition.
\begin{proposition}\label{Pro: k-opportunism}
We have $W_N^k\leq W_N^{k+1}$.
\end{proposition}
\begin{IEEEproof}
Similar to the proof of Proposition~\ref{Pro: 2}, we use induction on the number of requests.
For $N=1$, the claim follows immediately from the fact that both of the waiting times are equal to $\max_{m\leq M}\{T_{m1}\}$.
We now assume that the claim is true for $N-1$.
This means that $W_{M,N-1}^{k}\leq W_{M,N-1}^{k+1}$.
Since the number of links is irrelevant, this has to be true for every $m\leq M$, and we have $W_{m,N-1}^{k}\leq W_{m,N-1}^{k+1}$.
The proof proceeds as follows:
\begin{align*}
    W_N^k &= \max_{m\leq \min\{M, i+k-1\}} \{W_{m,N-1}^k+T_{mN}\}\\
    &\overset{(i)}{\leq}  \max_{m\leq \min\{M, i+k\}} \{W_{m,N-1}^k+T_{mN}\}\\
    &\overset{(ii)}{\leq}\max_{m\leq \min\{M, i+k\}} \{W_{m,N-1}^{k+1}+T_{mN}\} = W_N^{k+1}
\end{align*}
in which (i) holds since the set over which we take the maximum is enlarged, and (ii) is based on the induction hypothesis.
\end{IEEEproof}

Although $W_N$ is itself lower than $W^{\textsc{up}}_N$, opportunism has a lower bound and cannot arbitrarily decrease the total waiting time.
Indeed, a simple lower bound $W^{\textsc{low}}_N$ for $W_N$ can be achieved as follows.
\begin{proposition}\label{Pro: norm one}
The opportunistic waiting time is bounded from below, and we have $W^{\textsc{low}}_N = \max_{i\leq M}\sum_{j\leq N}T_{ij} \leq W_N$.
\end{proposition}
\begin{IEEEproof}
If we replace $\max_{m\leq i} \{W_{m,j-1}+T_{mj}\}$ with $W_{i,j-1}+T_{ij}$ in the definition of $W_{ij}$ in Equation~\eqref{EQ: opportunism}, this will not increase the value of $W_{MN}$, since we shrink the set we take the maximum on.
We therefore define $\tilde{W}_{ij}$ as follows:
\begin{equation}\label{eq: W_l}
    \tilde{W}_{ij} = \begin{cases}
    T_{ij} & j=1 \\
    \tilde{W}_{i,j-1}+T_{ij} & j > 1
\end{cases}
\end{equation}
We define $W^{\textsc{low}}_N = \tilde{W}_{MN}$.
Based on the definition above, we have $\tilde{W}_{MN}= \max_{i\leq M}\sum_{j\leq N}T_{ij}$.
Since this procedure cannot increase the waiting time, we have $\tilde{W}_{MN}\leq W_{MN}$, which means $W^{\textsc{low}}_N \leq W_N$.
\end{IEEEproof}

Just as the $k$-opportunistic settings form a spectrum between 1-opportunism and non-opportunism, i.e., $W_N$ and $W^{\textsc{up}}_N$, one can wonder if there is such a spectrum between $W^{\textsc{low}}_N$ and $W_N$.
It turns out that such a spectrum exists, and $W^{\textsc{low}}_N$ and $W_N$ are the two extremes of it.
We can demonstrate this spectrum with the following new definition:
\begin{equation}\label{EQ: memory}
\hat{W}_{ij}^{r} = \begin{cases}
\max_{\max\{0, i-r\}\leq k\leq i} T_{kj} & j=1 \\
\max_{\max\{0, i-r\}\leq k\leq i} \{\hat{W}_{k,j-1}^r+T_{kj}\} & j > 1
\end{cases}
\end{equation}
We now define $\hat{W}_N^{r}=\max_{i\leq M}\hat{W}^r_{iN}$.
Note that for $r=0$ we have $\hat{W}_N^{0}=W_{\textsc{low}}$, and for $r=M$ we have $\hat{W}_N^{M}=W_N$, which shows that $W_{\textsc{low}}$ and $W_N$ are members of the same family, with parameter $r$.
Since this parameter specifies how many links we go back when searching for the maximum value, we call it the search depth.
\par
We now show that, similar to $k$-opportunism, the new family of $\hat{W}_N^r$ form an ascending order based on $r$.
\begin{proposition}\label{Pro: memory}
We have $\hat{W}_N^r\leq \hat{W}_N^{r+1}$.
\end{proposition}
\begin{IEEEproof}
We use induction on the number of requests, $N$, to show that for every $i$ we have $\hat{W}_{iN}^r\leq \hat{W}_{iN}^{r+1}$.
This way, since $\hat{W}_N^r=\max_{i\leq M}\hat{W}_{iN}^r$ and $\hat{W}_N^{r+1}=\max_{i\leq M}\hat{W}_{iN}^{r+1}$, the claim is proved.
First assume that $N=1$.
In this case, the induction claim follows from the fact that taking maximum over a larger group does not decrease the maximum value.
Therefore, since when we go back $r+1$ links to search for maximum in calculating $\hat{W}_{i1}^{r+1}$, while we go back only $r$ links when calculating $\hat{W}_{i1}^r$, the claim is proved.
\par
Let us now assume that the induction claim is true for $N-1$.
We have:
\begin{align*}
    \hat{W}_{iN}^{r} &= \max_{\max\{0, i-r\}\leq k\leq i} \{\hat{W}_{k,N-1}^r+T_{kN}\}\\
    &\leq \max_{\max\{0, i-(r+1)\}\leq k\leq i} \{\hat{W}_{k,N-1}^r+T_{kN}\} \\
    &\overset{(i)}{\leq} \max_{\max\{0, i-(r+1)\}\leq k\leq i} \{\hat{W}_{k,N-1}^{r+1}+T_{kN}\} = \hat{W}_{iN}^{r+1}
\end{align*}
Inequality (i) follows from the induction hypothesis.
\end{IEEEproof}

We have now established two families of ``waiting times", and have shown that $W_N$ is in fact sandwiched between these two families.
From below, it is bounded by the different search depths, and from above it is bounded by degrees of opportunism.
Note that, in contrast to degrees of opportunism, different search depths do not have a real counterpart in the algorithm, and are mere mathematical tools to have a better understanding of the mathematical structure of the problem.
We can gather the results of this section in the following theorem.
\begin{theorem}\label{thm: spectrum}
In a line network, depths of search and degrees of opportunism sandwich the total waiting time in the opportunistic setting from below and above, respectively:
\begin{align*}
    \max_{j\leq N}\sum_{i\leq M}T_{ij} = &\hat{W}_N^0 \leq \hat{W}_N^1 \dots \leq \hat{W}_N^M \\
    &= W_N \\
    &= W_N^1 \leq W_N^2 \dots\leq W_N^M = \sum_{j\leq N}\max_{i\leq M} T_{ij}
\end{align*}
\end{theorem}

\subsection{Transmission Rate}\label{Sec: transmission rate}
A natural question to ask is the effect of opportunism on the transmission rate between $A$ and $B$.
The rate of interest is the number of requests sent from $A$ to $B$ per step.
To characterize this rate, we define the random variable $N_T$ to be the number of requests delivered from $A$ to $B$ up to time $T\in\mathbb N$.
Based on this definition, we define the transmission rate between $A$ and $B$ to be:
\begin{equation}\label{EQ: rate}
    R = \lim_{t\rightarrow\infty}\frac{E\{N_t\}}{t}
\end{equation}
The first thing we have to do is to prove such a limit exists.
In order to do so, we will use the following lemma~\cite{Fekete}.
\begin{lemma}\label{lem: Fekete}
(Fekete's Lemma) For every superadditive (subadditive) sequence $\{a_n\}$, the limit $\lim_{n\rightarrow\infty}\frac{a\{n\}}{n}$ is equal to $\sup \frac{a_n}{n}$ ($\inf \frac{a_n}{n}$).
\end{lemma}

To use this lemma, we need to prove the following proposition.
\begin{proposition}\label{Pro: superadditive}
The sequence $E\{N_t\}$ is superadditive.
That is, for two integers $t_1$ and $t_2$ we have $E\{N_{t_1+t_2}\}\geq E\{N_{t_1}\} + E\{N_{t_2}\}$.
\end{proposition}
\begin{IEEEproof}
Without loss of generality, let us assume $t_1\leq t_2$.
After $t_1$ time steps have passed, a total of $N_{t_1}$ requests have been sent to $B$.
Let us show the number of requests sent in the remaining $t_2$ time steps with $X$.
Then we have $N_{t_1+t_2} = N_{t_1} + X$, which means $E\{N_{t_1+t_2}\} = E\{N_{t_1}\} + E\{X\}$.
Now, note that after $t_1$ time steps, some of the links of the network are already established, and some of the pending requests are somewhere between $A$ and $B$.
Therefore, if we compare $X$ to a situation in which the network starts from scratch and runs for $t_2$ seconds, we can see that in expectation, $E\{X\}\geq E\{N_{t_2}\}$.
This leads us to $E\{N_{t_1+t_2}\} \geq E\{N_{t_1}\} + E\{N_{t_2}\}$, and finishes the proof.
\end{IEEEproof}

We can now prove the existence of $R$ as defined in Equation~\eqref{EQ: rate}.
\begin{proposition}\label{Pro: N limit exists}
The limit $R=\lim_{t\rightarrow\infty}\frac{E\{N_t\}}{t}$ exists and is finite.
\end{proposition}
\begin{IEEEproof}
The existence follows immediately from Lemma~\ref{lem: Fekete} and Proposition~\ref{Pro: superadditive}.
To show that $R$ is finite, note that for every $t\in\mathbb N$ we have $N_t\leq t$.
This is true because the only case in which we can have $N_t = t$ is when all of the links succeed consecutively in generation, up until time $t$, so that in each step a single request is able to use all of the links to reach $B$.
Therefore, no more than $t$ requests can be sent, since this is the best case.
We can thus see that for every $t\in \mathbb N$ we have $\frac{E\{N_t\}}{t}\leq 1$, which means $R\leq 1$.
\end{IEEEproof}

Although the convergence of $\frac{E\{N_t\}}{t}$ to a constant is desirable enough, it would be better if we could say more, specially about the dynamics of $\frac{N_t}{t}$.
In fact, using the standard bounded difference method~\cite{Steele}, we show that $\frac{N_t}{t}$ converges almost surely to $R$.
We need to introduce some more structure before proving this claim.
The following construction is a straightforward application of the procedure described in~\cite{Steele}.

We start by introducing a new group of random variables, based on an idea from~\cite{Delay-Line-Networks}.
Let us show the state of link $j\in\{1, \dots, M\}$ at time $i\in\{1, \dots, t\}$ by $X_{ji}\in\{0, 1\}$, where $X_{ji} = 1$ ($X_{ji} = 0$) means that link $j$ is generated (not generated) at time $i$.
Notice that $\{X_{ji}\}$ are not independent.
Based on these variables, let $\mathcal{F}_{ji}$ be the sigma-field generated by the $X_{ab}$ variables up to $a=j$ and $b=i$, i.e., $\mathcal{F}_{ji} = \sigma(X_{11}, X_{21}, \dots, X_{ji})$.
Note that these sigma-fields ``grow" based on a lexicographic order.

We define $d_{ji}=E\{N_t|\mathcal{F}_{ji}\} - E\{N_t|\mathcal{F}_{\alpha(j,i)}\}$, in which $\alpha(j, i)$ gives the previous $(j', i')$ in the lexicographic order, i.e., $\alpha(j, i) = (j-1, i)$ if $j > 1$ and $\alpha(1, i) = (M, i-1)$.
The purpose of this formality is to ensure the $d_{ji}$ grow based on the order of the $\mathcal{F}_{ji}$.
Note that we have $N_t = N_t(X_{11}, X_{21},\dots, X_{Mt})$, which means $E\{N_t|\mathcal{F}_{Mt}\} = N_t$.
This leads us to $\sum_{i,j}d_{ji} = N_t - E\{N_t\}$.

We now proceed to prove that the $\{d_{ji}\}$ are bounded.
Let us introduce a new set of variables $\{N^{(ji)}_t\}$.
For every $j$ and $i$, $N^{(ji)}_t$ is the same as $N_t$, except that it forgets the actual value of $X_{ji}$, and assumes $X_{ji}=1$.
That is, we have:
\begin{align}\label{EQ: N_t difference}
    N^{(ji)}_t&(X_{11}, X_{21}, \dots, X_{ji}, \dots X_{Mt})\nonumber\\
    &= N_t(X_{11}, X_{21}, \dots, 1, \dots X_{Mt})
\end{align}
The following lemma is essential to our path towards the convergence of $\frac{N_t}{t}$.
\begin{lemma}\label{lem: N_t difference is bounded}
For every $i$ and $j$ we have:
\begin{align*}
    |N_t&(X_{11}, X_{21}, \dots, X_{ji}, \dots X_{Mt})\\
    &- N^{(ji)}_t(X_{11}, X_{21}, \dots, X_{ji}, \dots X_{Mt})|\leq 1
\end{align*}
\end{lemma}
\begin{IEEEproof}
If $X_{ji}=1$, then based on Equation~\eqref{EQ: N_t difference} we have $N_t - N^{(ji)}_t = 0$.
If $X_{ji} = 0$, then the best case scenario is when all of the links are consecutively generated, except for link $j$ at time $i$.
In this case, flipping $X_{ji}$ to one allows one more request to pass, and increases $N_t$ by one.
Therefore, having a link active at a certain time can increase $N_t$ by at most one amount, which means $|N_t - N^{(ji)}_t|\leq 1$.
\end{IEEEproof}

We now use Lemma~\ref{lem: N_t difference is bounded} to bound $||d_{ji}||_\infty$.
This will allow us to leverage a version of Azuma's inequality~\cite{Steele}.
We start with showing that $E\{N_t|\mathcal{F}_{ji}\}\leq E\{N^{(ji)}_t|\mathcal{F}_{\alpha(j,i)}\}$.
Because of the discrete nature of the problem, we can think of $E\{N_t|\mathcal{F}_{ji}\}$ as a function of a tuple $(x_{11}, \dots, x_{\alpha(j, i)}, x_{ji})$, in which $x_{ab}$ is a realization of $X_{ab}$~\cite{Jacod-Protter}.
We can take each one of such tuples and change it to $(x_{11}, \dots, x_{\alpha(j, i)}, 1)$, which is in a bijective correspondence with the tuple $(x_{11}, \dots, x_{\alpha(j, i)})$.
In order to use this, we
name $G = E\{N_t|\mathcal{F}_{ji}\}$, $H = E\{N^{(ji)}_t|\mathcal{F}_{\alpha(j,i)}\}$, and $H' = E\{N^{(ji)}_t|\mathcal{F}_{ji}\}$.
This allows us to form a bridge between $G$ and $H$, as follows.
First, note that given the knowledge of $\mathcal{F}_{\alpha(j,i)}$, $N^{(ji)}_t$ is totally decoupled from $X_{ji}$, and the future state of the links is independent from this variable.
Therefore, we have:
\begin{align}
    H(x_{11}, \dots, x_{\alpha(j,i)}) = H'(x_{11}, \dots, x_{\alpha(j,i)}, x_{ji})
\end{align}
Second, based on the reasoning in the proof of Lemma~\ref{lem: N_t difference is bounded}, we see that:
\begin{align}\label{EQ: G and H}
    G(x_{11}, \dots, x_{\alpha(j, i)}, x_{ji})&\leq H'(x_{11}, \dots, x_{\alpha(j, i)}, x_{ji})
\end{align}
Since Equation~\eqref{EQ: G and H} holds for every tuple $(x_{11}, \dots, x_{\alpha(j, i)}, x_{ji})$, we conclude that $G\leq H$, which means:
\begin{align}\label{EQ: first}
    E\{N_t|\mathcal{F}_{ji}\}\leq E\{N^{(ji)}_t|\mathcal{F}_{\alpha(j,i)}\}.
\end{align}
Quite similarly, if we define $G' = E\{N_t|\mathcal{F}_{\alpha(j,i)}\}$, we have:
\begin{align}\label{EQ: G' and H'}
    &G'(x_{11}, \dots, x_{\alpha(j, i)})\leq H(x_{11}, \dots, x_{\alpha(j, i)})\nonumber\\
    &= H'(x_{11}, \dots, x_{\alpha(j, i)}, x_{ji})
\end{align}
which means:
\begin{align}\label{EQ: second}
    E\{N_t|\mathcal{F}_{\alpha(j,i)}\} \leq E\{N^{(ji)}_t|\mathcal{F}_{ji}\}
\end{align}
Combining Inequalities~\eqref{EQ: first} and~\eqref{EQ: second} we achieve:
\begin{align}\label{EQ: bounds for d_ji}
    E\{N_t-N^{(ji)}_t|\mathcal{F}_{ji}\} \leq d_{ji} \leq E\{N^{(ji)}_t-N_t|\mathcal{F}_{\alpha(j,i)}\}
\end{align}
Now, we know from Lemma~\ref{lem: N_t difference is bounded} that $|N_t-N^{(ji)}_t|\leq 1$.
We also know that conditional expectation cannot increase the $||.||_\infty$ norm.
Therefore, from Inequality~\eqref{EQ: bounds for d_ji} we conclude that for every $i$ and $j$ we have $||d_{ji}||_\infty\leq 1$.
We are now ready to prove our claim.
\begin{proposition}\label{Pro: N_t converges almost surely}
With probability one, $\frac{N_t}{t}$ converges to $R$.
\end{proposition}
\begin{IEEEproof}
Using a version of Azuma's inequality~\cite{Steele}, we can establish the following tail bound.
Note that we have replaced $\sum_{i,j} d_{ji}$ with $N_t - E\{N_t\}$:
\begin{align}
    P(|N_t - E\{N_t\}|\geq \lambda) \leq 2\exp\left(\frac{-\lambda^2}{2\sum_{i,j}||d_{ji}||^2_\infty}\right)
\end{align}
Using the fact that for every $i$ and $j$ we have $||d_{ji}||_\infty\leq 1$, we can change this inequality to:
\begin{align}\label{EQ: weaker Azuma}
    P(|N_t - E\{N_t\}|\geq \lambda) \leq 2\exp\left(\frac{-\lambda^2}{2Mt}\right)
\end{align}
For every $t\in\mathbb N$, and a fixed $0 < \epsilon\in\mathbb R$ let us define $\lambda_t = \sqrt{2Mt(\log t)(1+\epsilon)}$, and the corresponding event $A_t = \{|N_t - E\{N_t\}|\geq\lambda_t\}$.
Using Inequality~\eqref{EQ: weaker Azuma} we have:
\begin{align}
    \sum_{t}P(A_t)\leq \sum_{t}\frac{1}{t^{1+\epsilon}} < \infty.
\end{align}
Therefore, based on the Borel-Cantelli lemma we have $P(A_t, i.o.) = 0$.
This means that with probability one we have $N_t-E\{N_t\} = O(\sqrt{t\log t})$, or $\frac{N_t}{t}-\frac{E\{N_t\}}{t} = O(\sqrt{\frac{\log t}{t}})$.
Therefore, since $\frac{E\{N_t\}}{t}$ tends to $R$ as $t\rightarrow\infty$, we conclude that $\frac{N_t}{t}$ should also tend to $R$.
\end{IEEEproof}

Based on the analysis of the waiting time in Section~\ref{Sec:NReq}, we can readily obtain a lower and an upper bound for $R$.
To see this, first notice that based on the definition of $N_t$, we have the following equality:
\begin{align}\label{EQ: N based on W}
N_t = \max\{n\in\mathbb N \;|\; W_n\leq t\}.
\end{align}
Let us define $N^{\textsc{low}}_{t}$ and $N^{\textsc{up}}_{t}$ in a similar fashion:
\begin{align}\label{EQ: N up and low}
&N^{\textsc{up}}_t = \max\{n\in\mathbb N \;|\; W^{\textsc{up}}_n\leq t\}\\
&N^{\textsc{low}}_t = \max\{n\in\mathbb N \;|\; W^{\textsc{low}}_n\leq t\}
\end{align}
Now, since we know $W^{\textsc{low}}_n\leq W_n\leq W^{\textsc{up}}_n$, we can see that $N^{\textsc{low}}_t\geq N_t\geq N^{\textsc{up}}_t$.
This allows us to have the following proposition, which we just proved.
\begin{proposition}
Let us define (notice the reversal of the \textsc{low} and \textsc{up} subscripts):
\begin{align}\label{EQ: R up and low}
&R_{\textsc{low}}=\lim_{t\rightarrow\infty}\frac{E\{N^{\textsc{up}}_t\}}{t}\\
&R_{\textsc{up}}=\lim_{t\rightarrow\infty}\frac{E\{N^{\textsc{low}}_t\}}{t}
\end{align}
We then have $R_{\textsc{low}}\leq R\leq R_{\textsc{up}}$.
\end{proposition}

The same reasoning can be used to define the corresponding rates for the $k$-opportunistic waiting times, as well as the different search depths.
Thus, the spectrum shown in Theorem~\ref{thm: spectrum} can be directly turned into a spectrum of rates.
We do not repeat the process for brevity.

It turns out that there exists another interesting upper bound for $R$.
This upper bound stems from a slightly different definition for the transmission rate in the opportunistic setting.
Let us define:
\begin{align}\label{EQ: rate with w}
   \tilde{R} = \lim_{n\rightarrow\infty}\frac{n}{E\{W_n\}}.
\end{align}
Similar to Proposition~\ref{Pro: N limit exists}, one first has to show $\tilde{R}$ exists and is finite.
This can be done in the same manner, as can be seen from the next proposition.
\begin{proposition}
The limit $\tilde{R}=\lim_{t\rightarrow\infty}\frac{n}{E\{W_n\}}$ exists and is finite.
\end{proposition}
\begin{IEEEproof}
The procedure is very similar to that of Proposition~\ref{Pro: N limit exists}.
We first show that the sequence $E\{W_n\}$ is subadditive.
Assume two integers $n_1\leq n_2$, and let us start with $W_{n_1+n_2}$.
After the first $n_1$ requests have been sent, $W_{n_1}$ time steps have passed.
Let us show the remaining time steps until all $n_1+n_2$ requests have been sent with $Y$.
We thus have $W_{n_1+n_2} = W_{n_1} + Y$.
After $W_{n_1}$ time steps, some of the links are established and some requests are already travelling through the network.
Therefore, $E\{Y\}$ cannot be more than the case in which the network starts from scratch, leading to $E\{Y\}\leq E\{W_{n_2}\}$.
We thus conclude that $E\{W_{n_1+n_2}\}\leq E\{W_{n_1}\} + E\{W_{n_2}\}$.

The existence of $\tilde{R}$ follows immediately from Lemma~\ref{lem: Fekete}.
To see that $\tilde{R}$ is also finite, it is sufficient to note that for every $n\in \mathbb N$ we always have $n\leq W_n$.
This happens because the lowest $W_n$ can get is when all of the links succeed consecutively in generation, up until request $n$, so that in each step a single request is able to use all of the links to reach $B$.
Therefore, we always have $\frac{n}{E\{W_n\}}\leq 1$, which means $\tilde{R}$ is finite.
\end{IEEEproof}

Although quite similar in nature, simulations indicate that $R$ and $\tilde{R}$ are different, and we have $R\leq\tilde{R}$.
We do not have a proof for this yet, and postpone a demonstration of this curious point to the simulations in Section~\ref{Sec: simulations}.

\subsection{Further Investigation}
Let us get back to the matrix $D$, defined in Equation~\eqref{EQ: D}.
As explained in Section~\ref{Sec:NReq}, we can introduce different degrees of opportunism and searching depths when analyzing the waiting times.
There is a mathematical waiting time associated with every one of these cases.
Thinking of these waiting times as mere functions of matrices, it turns out that they have an interesting property.
Put simply, these waiting times act as norms for general matrices.
Furthermore, the $||.||_{\infty}$ matrix norm is a natural member of this family of norms, as we will show here.

To analyze the waiting time functions as norms, we need to define them in a different, more systematic manner.
The following definition introduces the building blocks of such a systematic approach.
\begin{definition}\label{def: theta}
For every $0\leq r\leq M$ and $1\leq k\leq M$, and for vectors $a,b\in\mathbb R^{M}$, we define the function $\Theta_{r}^{k}: \mathbb R^{M}\times\mathbb R^{M}\rightarrow\mathbb R^{M}$ as follows:
\begin{align*}
    \Theta_{r}^{k}(a,b)=(v_1, \dots, v_M)
\end{align*}
{\centering
  $ \displaystyle
    \begin{aligned} 
       v_i = \max_{\max\{0, i-r\}\leq j\leq\min\{i+k-1, M\}}\{|a_j|+|b_j|\}
    \end{aligned}
  $ 
\par}\vspace{3pt}
\end{definition}
Based on this definition, we have the following lemma.
\begin{lemma}\label{lem: theta}
For every $0\leq r\leq M$ and $1\leq k\leq M$, every $\alpha\in\mathbb R$, and for vectors $a,b\in\mathbb R^{M}$, the function $\Theta_{r}^{k}(a,b)$ satisfies the following two properties:
\begin{enumerate}
    \item If $\Theta_{r}^{k}(a, b) = 0$, then $a=0$ and $b=0$.
    \item $\Theta_{r}^{k}(\alpha a, \alpha b) =\Theta_{r}^{k}(-\alpha a, \alpha b) = \Theta_{r}^{k}(\alpha a, -\alpha b) =  |\alpha|\Theta_{r}^{k}(a,b)$.
\end{enumerate}
\end{lemma}
\begin{IEEEproof}
We first prove the first property.
Based on Definition~\ref{def: theta}, $\Theta_{r}^{k}(a, b) = 0$ means that for every $i$ and every $j$ between $\max\{0,i-r\}$ and $\min\{i+k-1, M\}$, we have $|a_j|+|b_j|\leq 0$, which means $a_j = b_j = 0$.
Since we can cover all $1\leq j\leq M$ by sliding $i$, we conclude that $a=b=0$.
\par
We now prove the second property.
Assuming $\Theta_{r}^{k}(a, b)=(v_1, \dots, v_M)$ and $\Theta_{r}^{k}(\alpha a, \alpha b)=(u_1, \dots, u_M)$, for every $i$ we have:
\begin{align*}
    u_i &= \max_{\max\{0, i-r\}\leq j\leq\min\{i+k-1, M\}}\{|\alpha a_j|+|\alpha b_j|\} \\
    &= \max_{\max\{0, i-r\}\leq j\leq\min\{i+k-1, M\}}\{|\alpha|(|a_j|+|b_j|)\} \\
    &= |\alpha|\max_{\max\{0, i-r\}\leq j\leq\min\{i+k-1, M\}}\{|a_j|+|b_j|\} \\
    &= |\alpha|v_i
\end{align*}
We thus conclude that $\Theta_{r}^{k}(\alpha a, \alpha b)=|\alpha|\Theta_{r}^{k}(a,b)$.
The other equalities are proved in the exact same way.
\end{IEEEproof}

Let us show the standard basis vectors for $\mathbb R^M$ with $\{e_i\}_{1\leq i\leq M}$.
That is, $e_i$ is the vector with a one in its $i$-th coordinate, and a zero in the rest.
For a given matrix $D$, we define the vectors $D_i$ as follows:
\begin{align}\label{eq: D_i}
    D_i = \Theta_{r}^{k}(D_{i-1}, e_{i}^{T}D)
\end{align}
With the initial vector $D_0=(0, \dots, 0)$.
We are now ready for the following theorem.
\begin{theorem}
For any matrix $D\in\mathbb R^{N\times M}$, and every $0\leq r\leq M$ and $1\leq k\leq M$, the following function is a norm:
\begin{equation}
    \Lambda_{r}^{k}(D)=||D_N||_\infty
\end{equation}
In which $D_N$ is based on Equation~\eqref{eq: D_i}.
\end{theorem}
\begin{IEEEproof}
We prove this theorem by checking the three conditions a real function on matrices should satisfy in order to be a matrix norm.
First, we show that if $\Lambda_{r}^{k}(D)=0$, then $D$ is the zero matrix.
Since $\Lambda_{r}^{k}(D)=0$, then $||D_N||_\infty = 0$.
This means that $D_N=0$.
Now, based on Lemma~\ref{lem: theta}, we have $e_N^{T}D=0$ and $D_{N-1}=0$.
The first equality means that the last row of $D$ is equal to the zero vector, and the second equality can be used in a recursive manner (based on Definition~\ref{def: theta}).
Each step $i$ reveals that the $i$-th row of $D$ is the zero vector.
Therefore, we have $D=0$.
\par
We now show that, for any $\alpha\in\mathbb R$, we have $\Lambda_{r}^{k}(\alpha D) = |\alpha|\Lambda_{r}^{k}(D)$.
Let us denote $\hat{D} = \alpha D$.
By using induction, we show that for every $i$ we have $\hat{D}_i = |\alpha|D_i$.
This way, we have $\hat{D}_N=|\alpha| D_N$, which directly leads to $\Lambda_{r}^{k}(\alpha D) = |\alpha|\Lambda_{r}^{k}(D)$.
For $i=1$, based on Lemma~\ref{lem: theta} we have $\hat{D}_1=\Theta_r^{k}(0, \alpha e_1^{T}D) = |\alpha|\Theta_r^{k}(0, e_1^{T}D) = |\alpha| D_1$.
Now, assuming the claim is true for $i$, we have:
\begin{align*}
    \hat{D}_{i+1} &= \Theta_r^{k}(\hat{D}_i, \alpha e_{i+1}^{T}D) = \Theta_r^{k}(|\alpha|D_i, \alpha e_{i+1}^{T}D)\\
    &= |\alpha|\Theta_r^{k}(D_i, e_{i+1}^{T}D) = |\alpha|D_{i+1}
\end{align*}
in which we have used the induction hypothesis and Lemma~\ref{lem: theta}.

Finally, we have to prove that $\Lambda_r^{k}$ satisfies the triangle inequality.
We will do so by showing that for the two matrices $C$ and $D$ we have $\Lambda_r^{k}(C+D)\leq \Lambda_r^{k}(C)+\Lambda_r^{k}(D)$.
Let us name $G = C+D$.
Also, let us show the $j$-th element of a vector $C_i$ with $(C_i)_j$.
In order to do prove the triangle inequality, we first use induction to prove that for every $i$ and $j$ we have $(G_i)_j \leq (C_i)_j + (D_i)_j$.
Once we have proved this, we have:
\begin{align*}
    \Lambda_r^{k}(G) &= ||G_N||_\infty = \max_{j}\{(G_N)_j\} \leq \max_{j}\{(C_N)_j + (D_N)_j\} \\
    &\leq \max_{j}\{(C_N)_j)\} + \max_{j}\{(D_N)_j\} \\
    &= \Lambda_r^{k}(C) + \Lambda_r^{k}(D)
\end{align*}
Which is the triangle inequality.
Now, in order to prove our claim based on induction, we first start with the base case of $i=1$.
Let us show the element in the $i$-th row and the $j$-th column of a matrix $C$ with $C_{ij}$.
We have:
\begin{align*}
    (G_1)_j &= \max_{\max\{0, j-r\}\leq s\leq\min\{j+k-1, M\}} \{|G_{1s}|\} \\
    &= \max_{\max\{0, j-r\}\leq s\leq \min\{j+k-1, M\}} \{|C_{1s} + D_{1s}|\} \\
    &\leq \max_{\max\{0, j-r\}\leq s\leq \min\{j+k-1, M\}} \{|C_{1s}| + |D_{1s}|\} \\
    &\leq \max_{\max\{0, j-r\}\leq s\leq \min\{j+k-1, M\}} \{|C_{1s}|\}\\
    &+ \max_{\max\{0, j-r\}\leq s\leq \min\{j+k-1, M\}} \{|D_{1s}|\} \\
    &= (C_1)_j + (D_1)_j
\end{align*}
Now we assume that our claim is true for $i-1$.
We have:
\begin{align*}
    (G_i)_j &= \max_{\max\{0, j-r\}\leq s\leq\min\{j+k-1, M\}} \{|(G_{i-1})_s| + |G_{is}|\} \\
    &= \max_{\max\{0, j-r\}\leq s\leq\min\{j+k-1, M\}} \{|(G_{i-1})_s| + |C_{is} + D_{is}|\} \\
    &\leq \max_{\max\{0, j-r\}\leq s\leq\min\{j+k-1, M\}} \{(|G_{i-1})_s| + |C_{is}| + |D_{is}|\} \\
    &\overset{(i)}{\leq} \max_{\max\{0, j-r\}\leq s\leq\min\{j+k-1, M\}} \{|(C_{i-1})_s| + |(D_{i-1})_s|\\
    &+ |C_{is}| + |D_{is}|\} \\
    &\leq \max_{\max\{0, j-r\}\leq s\leq\min\{j+k-1, M\}} \{|(C_{i-1})_s| + |C_{is}|\} \\ &+\max_{\max\{0, j-r\}\leq s\leq\min\{j+k-1, M\}} \{|(D_{i-1})_s| + |D_{is}|\} \\
    &= (C_i)_j + (D_i)_j
\end{align*}
This finishes our proof.
We have thus proved that the function $\Lambda_r^k$ is indeed a matrix norm.
\end{IEEEproof}

\section{Simulations}\label{Sec: simulations}
We will demonstrate our experimental results in this section.
These results can be divided to two main categories, each serving its own purpose.
We will first present simulation results for the line network analyzed in Section~\ref{Sec: analysis}, in which we demonstrate the proved claims numerically.
Second, using the simulator we have built, we present the effect of opportunism on a more complex network, under different design choices.
Our simulation setup in the second part is much more general than our analysis setup in~\ref{Sec: analysis}.
For instance, we use the grid topology for simulation, which automatically covers many complex routing scenarios.
We have also relaxed the assumptions of the analysis section in the second part of the simulations.

\subsection{Numerical Simulations for Line Networks}
We start with verifying our analytical results.
The analysis in Section~\ref{Sec: analysis} has three main themes.
The first is the fact that opportunism induces a spectrum of choices and corresponding waiting times.
The main point here is that opportunism decreases the total waiting time.
We postpone demonstrating these to Section~\ref{Sec: eval}, in which we use a much more complex network.
The second theme is about the asymptotic behavior of the transmission rate, whether in expectation or as a random variable.
Finally, the last theme is about bounding the transmission rate from above and below.
In the numerical simulations of this section, we focus on the transmission rate, and demonstrate the second and third themes of our analysis.

\begin{figure}[t]
\centerline{\includegraphics[width=1\columnwidth]{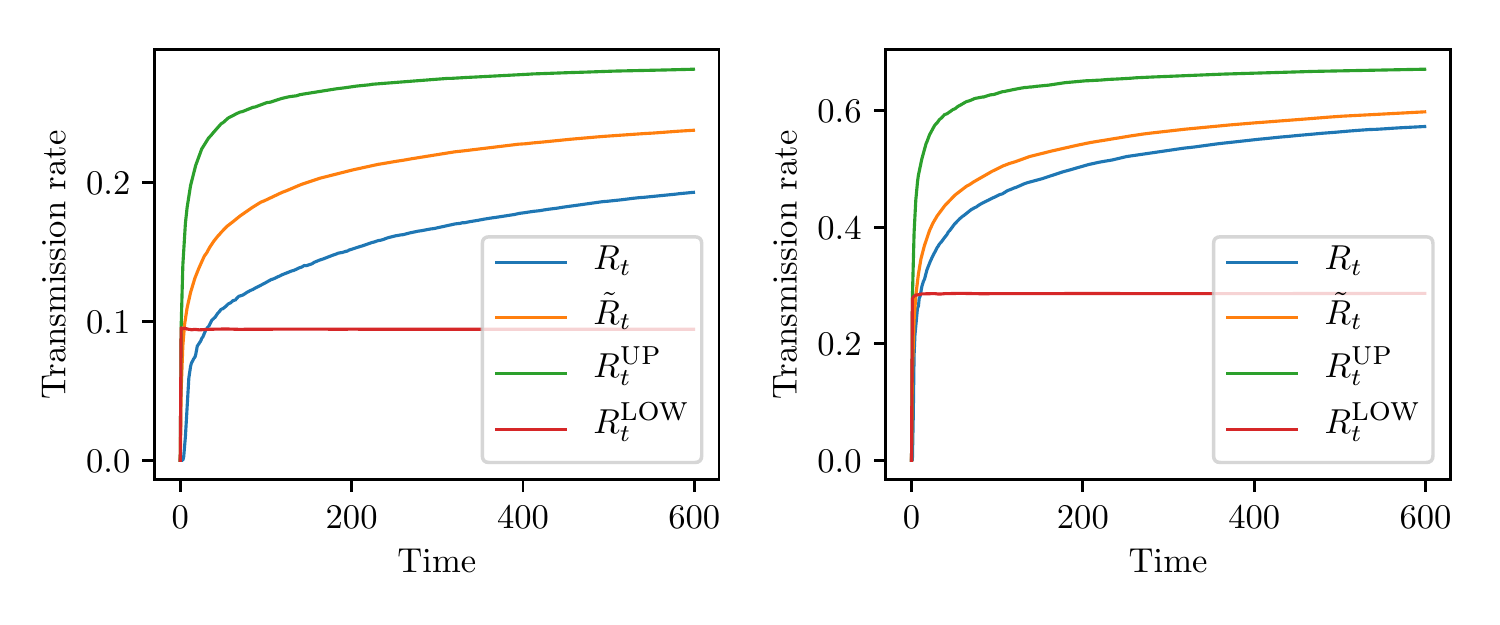}}
\caption{Transmission rate and its different bounds. In the figures on the left and the right we have $p_{\textsc{gen}}=0.3$ and $p_{\textsc{gen}}=0.7$, respectively.}
\label{fig: rates}
\end{figure}

Figure~\ref{fig: rates} shows the transmission rate
when we have $M=20$, for two values of $p_{\textsc{gen}}$.
Here, we have $R_t = \frac{E\{N_t\}}{t}$, $R^{\textsc{up}}_t = \frac{E\{N^{\textsc{low}}_t\}}{t}$, $R^{\textsc{low}}_t = \frac{E\{N^{\textsc{up}}_t\}}{t}$, and $\tilde{R}_t = \frac{t}{E\{W_t\}}$.
We can see that for every $t$, $R_t$ is more than $R^{\textsc{low}}_t$.
Now, we know from Equations~\eqref{EQ: N up and low} and~\eqref{EQ: R up and low} that $R^{\textsc{low}}_t$ corresponds to non-opportunism,
which means that in a line network, opportunism increases the transmission rate when compared to non-opportunism.
As we mentioned in Section~\ref{Sec: analysis}, $\tilde{R}_t$ seems to act as an upper bound to $R_t$, and we can see this as well in Figure~\ref{fig: rates}. 

\begin{figure}[htbp]
\centerline{\includegraphics[width=1\columnwidth]{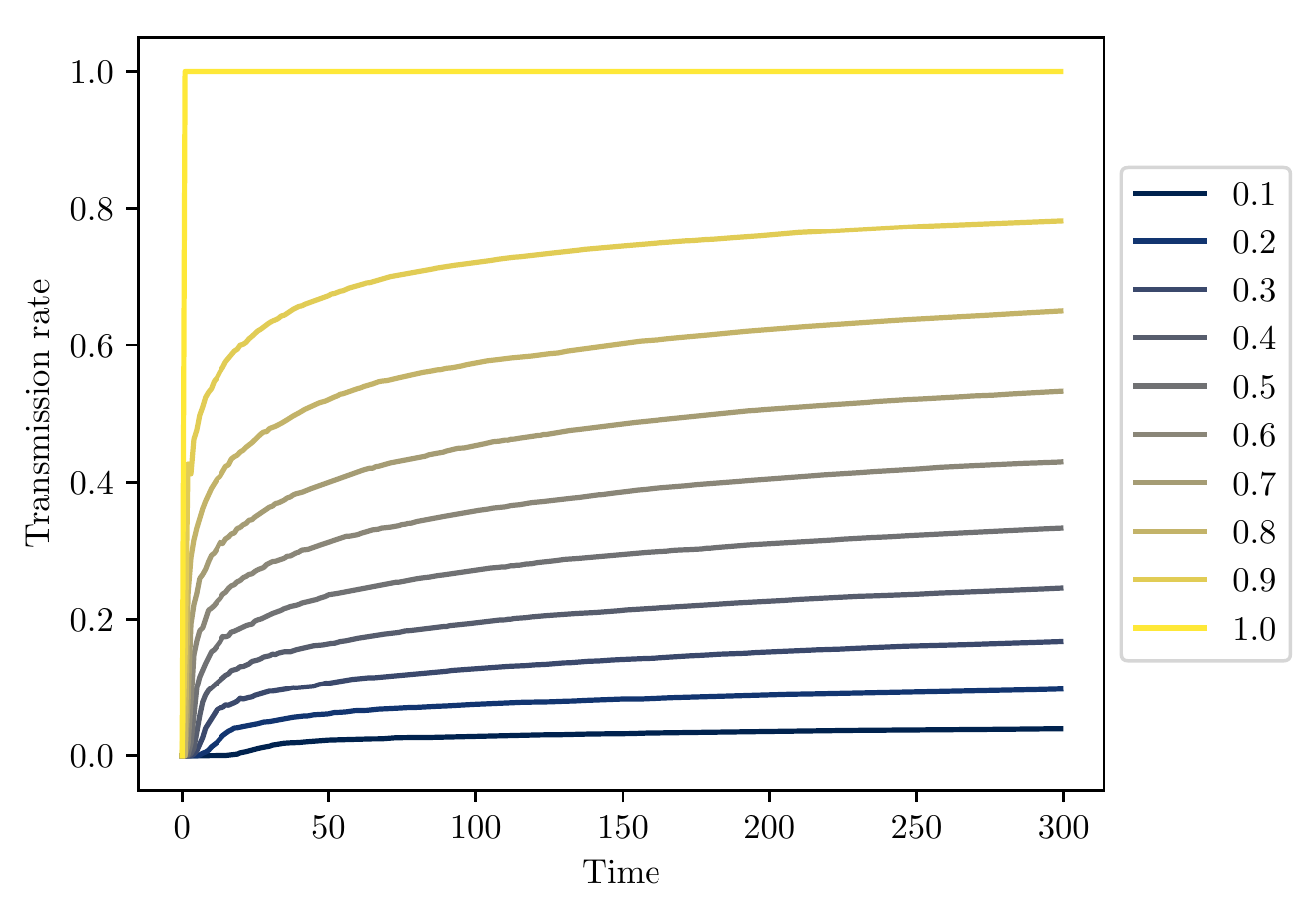}}
\caption{Transmission rate for different values of $p_{\textsc{gen}}$.
These values are specified in the legend. We have $M = 20$.}
\label{fig: rates-Pvar}
\end{figure}

\begin{figure}[htbp]
\centerline{\includegraphics[width=1\columnwidth]{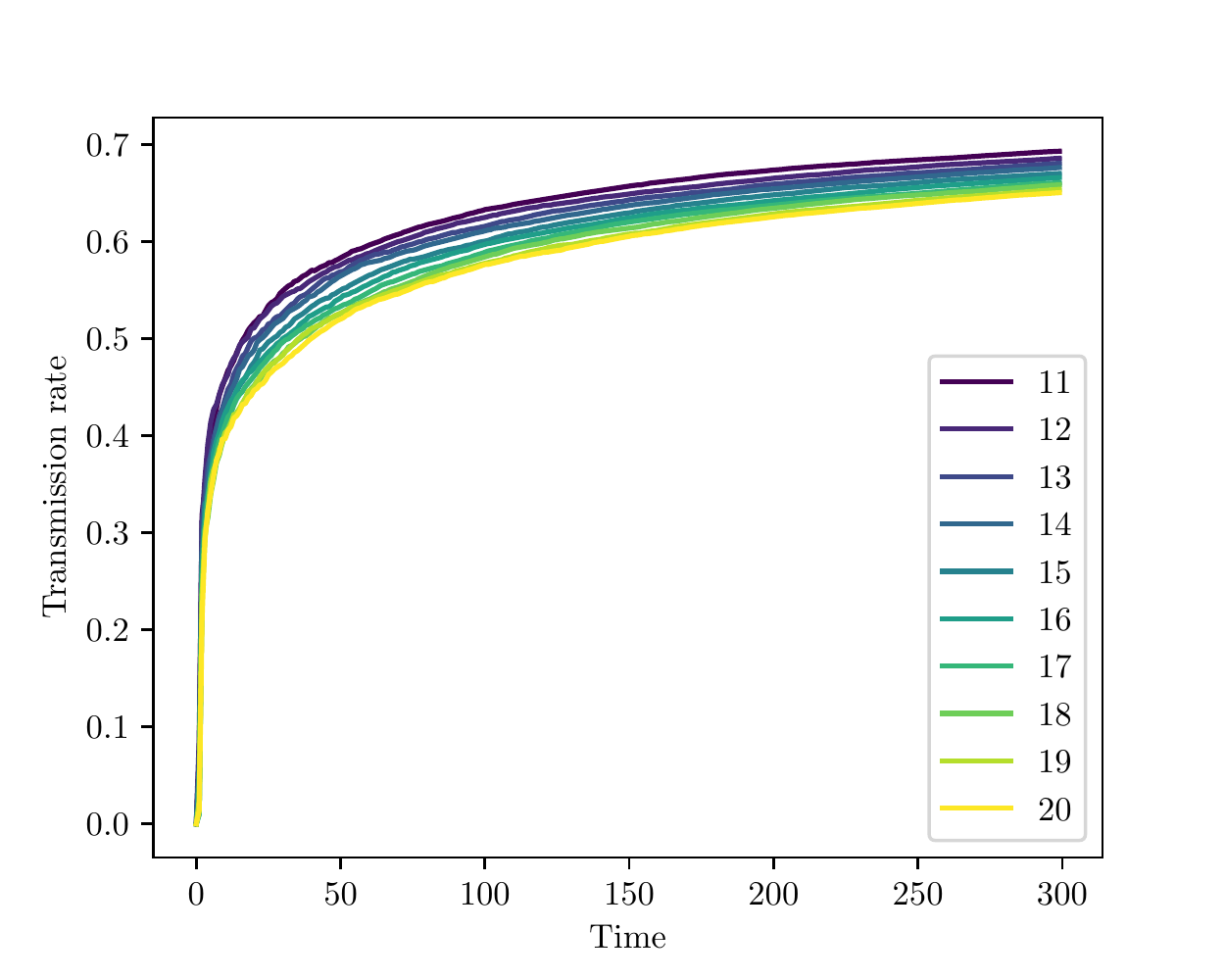}}
\caption{Transmission rate for different values of $M$.
These values are specified in the legend. We have $p_{\textsc{gen}} = 0.8$.}
\label{fig: rates-Mvar}
\end{figure}

We now demonstrate the behavior of $R_t$ based on the two main variables of the setting in our analysis, which are $p_{\textsc{gen}}$ and $M$.
We first fix the size of the network at $M=20$, and plot the transmission rate $R_t$ for different values of $p_{\textsc{gen}}$.
The result is shown in Figure~\ref{fig: rates-Pvar}, in which the value of $p_{\textsc{gen}}$ ranges from $0.1$ to $1$.
Similarly, we can fix $p_{\textsc{gen}} = 0.8$, and plot the transmission rate for different network sizes.
The results are demonstrated in Figure~\ref{fig: rates-Mvar}, in which $M$ ranges from $11$ to $20$.

\begin{figure}[htbp]
\centerline{\includegraphics[width=1\columnwidth]{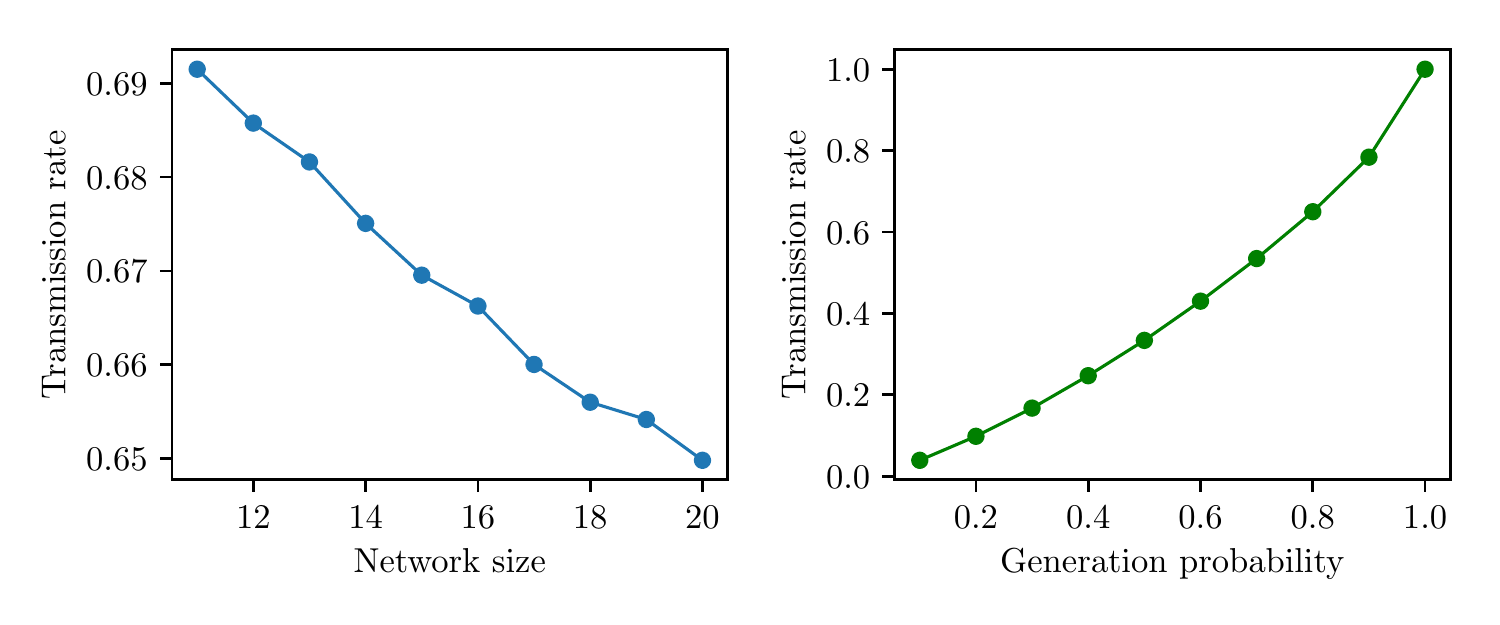}}
\caption{Transmission rate based on $M$ and $p_{\textsc{gen}}$. The value of $p_{\textsc{gen}}$ in the figure on the left is $0.8$, and the value of $M$ in the figure on the right is $20$.}
\label{fig: rate on M and p}
\end{figure}

These figures suggest that $M$ has a linear effect on the transmission rate, while $p_{\textsc{gen}}$ has a nonlinear one.
We can verify these observations by plotting the rate based on these variables, and not based on the time.
To do this, we select a time in which the rate has safely converged, and plot the rate in that time based on varying $M$ and $p_{\textsc{gen}}$ values.
The resulting plots are shown in Figure~\ref{fig: rate on M and p}.
This figure also shows that the rate is not very sensitive to the length of the line.
This is intuitively correct, because once enough requests have entered the network, the rate with which they reach the destination is not much related to the length of the network, and the generation probability is the main factor.

\begin{figure}[htbp]
\centerline{\includegraphics[width=1\columnwidth]{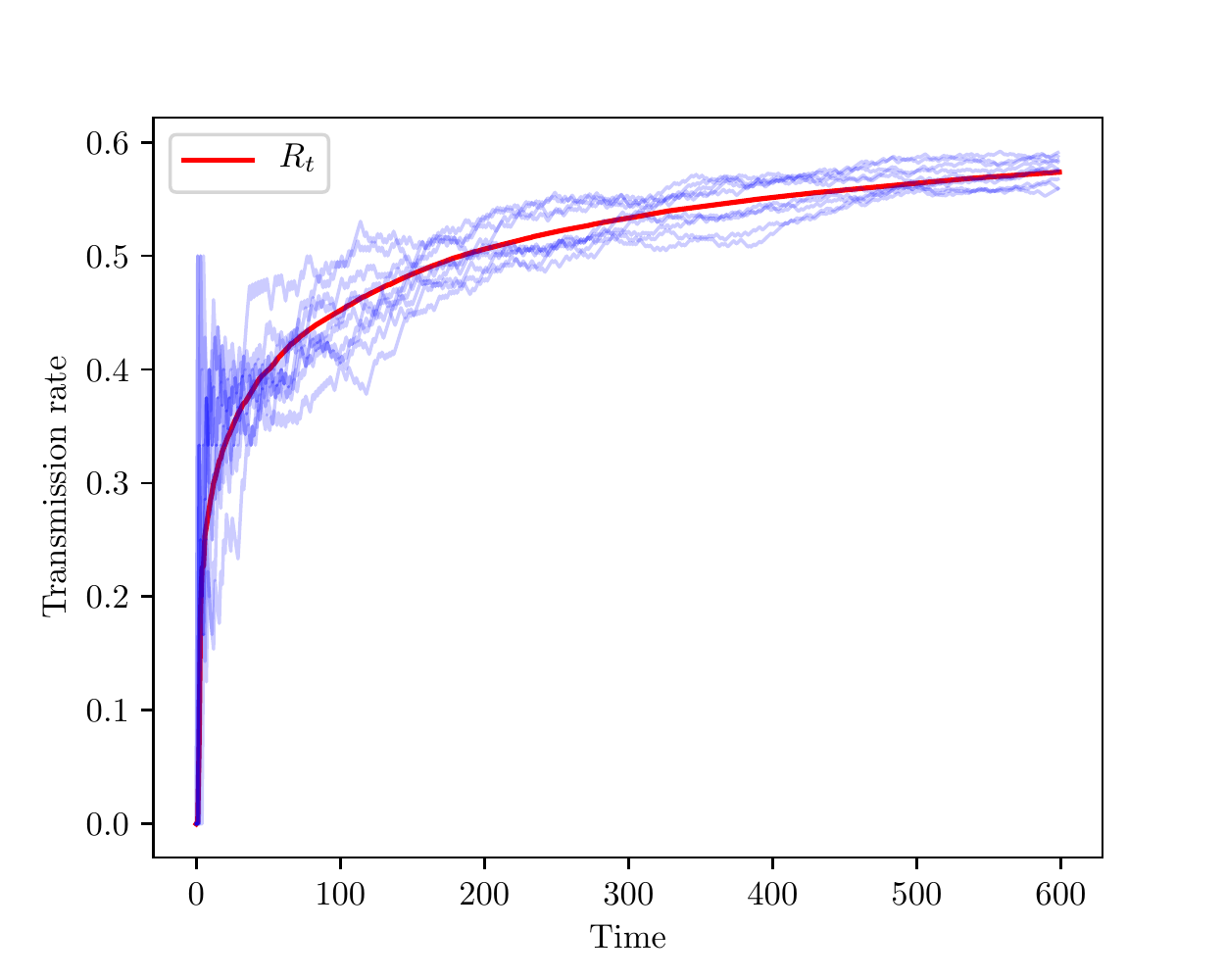}}
\caption{The stochastic convergence of the transmission rate. The values of $M$ and $p_{\textsc{gen}}$ are $20$ and $0.7$, respectively.}
\label{fig: stochastic rate}
\end{figure}

We finally demonstrate the stochastic behavior of the transmission rate.
As we proved in Section~\ref{Sec: analysis}, the transmission rate almost surely converges to the limit of its expectation.
Figure~\ref{fig: stochastic rate} shows this convergence for a number of simulated runs.

\subsection{QNS-Lite}
\begin{figure}[b]
\centerline{\includegraphics[width=1\columnwidth]{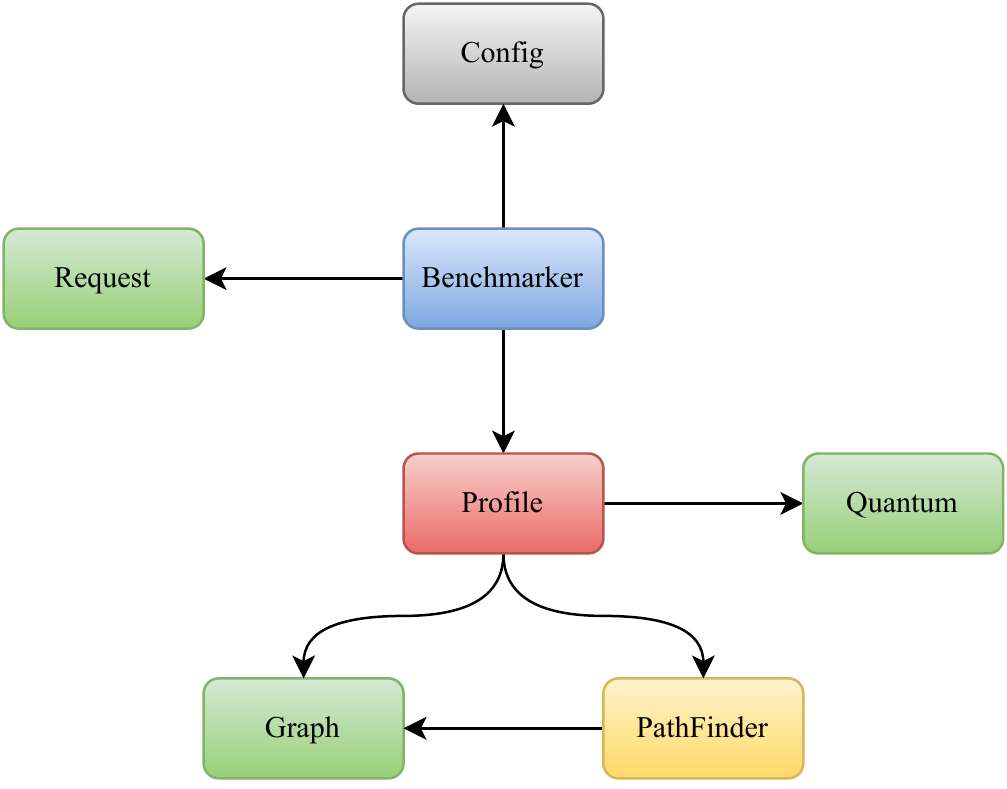}}
\caption{The modular structure of our simulator.}
\label{fig: modular simulator}
\end{figure}

Before presenting the evaluation results, we first provide a brief explanation of the simulator we developed during our research on opportunism, \emph{QNS-Lite}~\cite{QNS-Lite}.
We do so by explaining the logical, modular structure of the simulator, shown in Figure~\ref{fig: modular simulator}.
We go through the modules one by one:
\begin{itemize}
    \item \textbf{Graph:} This module provides the necessary abstractions (e.g., nodes and links) for the topology of the network.
    \item \textbf{Quantum:} The Quantum module captures the quantum behaviour of the system.
    This is done through a number of mathematical abstractions that represent the physical behaviour of the system.
    For instance, every generation attempt is modeled by a biased coin toss.
    \item \textbf{Request:} The necessary components of a simulated request are packed in the Request module.
    \item \textbf{Config:} This module contains the generic parameters of the system (e.g., $p_{\textsc{gen}}$), and allows the designer to specify these parameters in one place.
    \item \textbf{PathFinder:} The functionality of the PathFinder module is to find a number of (recovery) paths for the requests, based on different design choices.
    Different routing algorithms can have different PathFinder modules.
    \item \textbf{Profile:} This is the main module of the system.
    Here, a network topology and a specific routing algorithm are determined, and the simulation starts to run.
    Each combination of different networks and routing algorithms gives a different profile, and this module enables the common functionalities (e.g., forwarding) required for such algorithms to run.
    \item \textbf{Benchmarker:} The Benchmarker module handles the creation of the requests, and then runs the profile module.
    Furthermore, the metrics of the simulation are gathered by this module.
    
\end{itemize}


Following the existing literature on quantum routing (e.g.,~\cite{DistRoutQuantumInternet, ConcurrentRouting}), we have adopted the practice of dividing time into time slots, and furthermore, based on the structure explained in Section~\ref{Sec: quantum routing}, each time slot is divided to several stages.
Because of our focus on demonstrating the essential superiority of opportunism, we have kept some aspects of the design simple, and improving these aspects is a part of our future work.
This being said, most of the design choices required for creating and fine-tuning a quantum routing algorithm are crafted into the simulator, hence making it a suitable tool for research and development.
Finally, to have a more realistic evaluation, we have relaxed the assumption that swapping takes no time.
At the end of each time slot, we allow every request a single swapping opportunity.
We use an ordinary chain swapping from the source to the destination.
Note that, although there are better swapping methods, this choice does not harm our goal of demonstrating the superiority of opportunism.

\subsection{Evaluation}\label{Sec: eval}
In order to provide and explain our experimental results, we first explain our setup, metrics, and methodology, as follows.
\subsubsection{Setup}
In our setup, the source and the destination of each request are randomly selected, based on a uniform distribution on the nodes.
Once the requests are generated, the runtime of the simulator starts, and it finishes when all of the requests have reached their destinations.
This can be thought of as giving an impulse as the input to a system and measuring its impulse response.
All of the metrics are measured during the runtime.

\par
Each configuration of interest for this paper can be shown as the tuple $(p_\textsc{gen}, p_\textsc{swap}, L, N, M, k)$, in which $p_\textsc{gen}$ is the generation success probability, $p_\textsc{swap}$ is the swapping success probability, $L$ is the lifetime of a link, $N$ is the number of requests, $M$ is a parameter determining the size of the network, and $k$ is the opportunism degree.
In a grid network, the nodes are arranged in an $M\times M$ mesh.

\subsubsection{Metrics}
The main metric for our evaluations is the average total waiting time for requests, as explained in Section~\ref{Sec: analysis}.
The lower this metric becomes, the sooner (on average) the requests reach their destinations, and the higher the average throughput of the network.
In addition to this metric, we briefly evaluate the average link waiting time.
The lower this metric gets, the more efficient resource consumption becomes.
We only demonstrate that in all of the implemented algorithms, opportunism helps reduce these metrics, and we do not provide a full comparison of the algorithms.
\subsubsection{Methodology}
For every $(p_\textsc{gen}, p_\textsc{swap}, L, N, M, k)$ configuration, we first generate $N$ requests in the network and then run each algorithm 100 times.
We measure each metric in every runtime and calculate its average across the 100 iterations.
Then, we repeat this procedure another 50 times, each time with a different set of requests.
Each one of these outer iterations produces an average total waiting time for its corresponding $N$ requests, and an average link waiting time for the links.
We calculate the average of these average values, and then compare them across the different algorithms we have implemented.
This allows us to assess the essence of the overall effect of opportunism on the performance of a given network, by including many groups of requests in many scenarios.

\subsection{Results}
We have implemented three state-of-the-art algorithms alongside their opportunistic versions.
These algorithms are Modified Greedy (\textbf{MG})~\cite{DistRoutQuantumInternet}, Nonoblivious Local (\textbf{NL}, we have chosen this name for this algorithm for ease of use)~\cite{RoutingEntanglementInternet}, and QPASS (\textbf{QP})~\cite{ConcurrentRouting}.
It is worth mentioning that our version of the QP protocol is slightly modified, and to be more exact, it is a combination of the QPASS and QCAST algorithms from~\cite{ConcurrentRouting}.
Furthermore, the original QP algorithm in~\cite{ConcurrentRouting} uses a novel metric, \emph{EXT}, when finding paths for requests in arbitrary networks.
Due to the complete homogeneity of the grid topology, the EXT metric is equivalent to the hop distance, in terms of the preference it induces on different paths.
Thus we have used the ordinary hop distance for our path selection, in all of the three algorithms.
In all of the figures, NOPP and OPP correspond to the non-opportunistic and opportunistic approaches, respectively.

\par
\begin{figure}[t]
\centerline{\includegraphics[width=1\columnwidth]{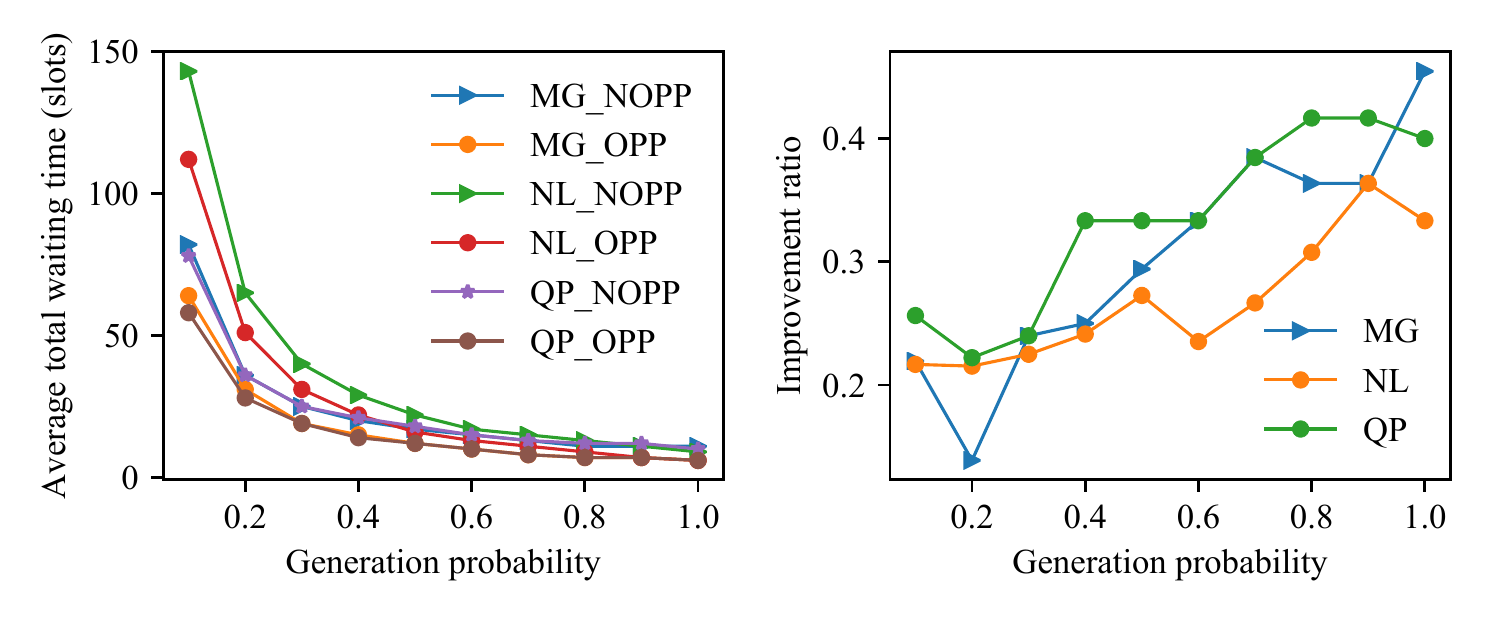}}
\caption{Average total waiting time in a grid network, with the fixed scenario $M=5$, $N=20$, $L=30$, $k=1$, and $p_\textsc{swap}=1$. The figure on the right shows the ratio of the improvement caused by opportunism to the non-opportunistic approach.}
\label{fig: results1}
\end{figure}

\textbf{Opportunism is Better.} We start by fixing $M = 5$, $N = 20$, $L = 30$, $k=1$, and $p_\textsc{swap} = 1$, and compare the average total waiting time based on $p_\textsc{gen}$.
Fig.~\ref{fig: results1} shows the result of this comparison.
The figure on the right shows the ratio of the improvement caused by opportunism to the non-opportunistic approach, which is obtained by dividing their difference by the non-opportunistic values.
It is clear that opportunism has improved the performance of the network by almost 30\%, and up to 45\%.
Furthermore, it can be seen that the QP algorithm provides better performance in this setup.

\par
\begin{figure}[htbp]
\centerline{\includegraphics[width=1\columnwidth]{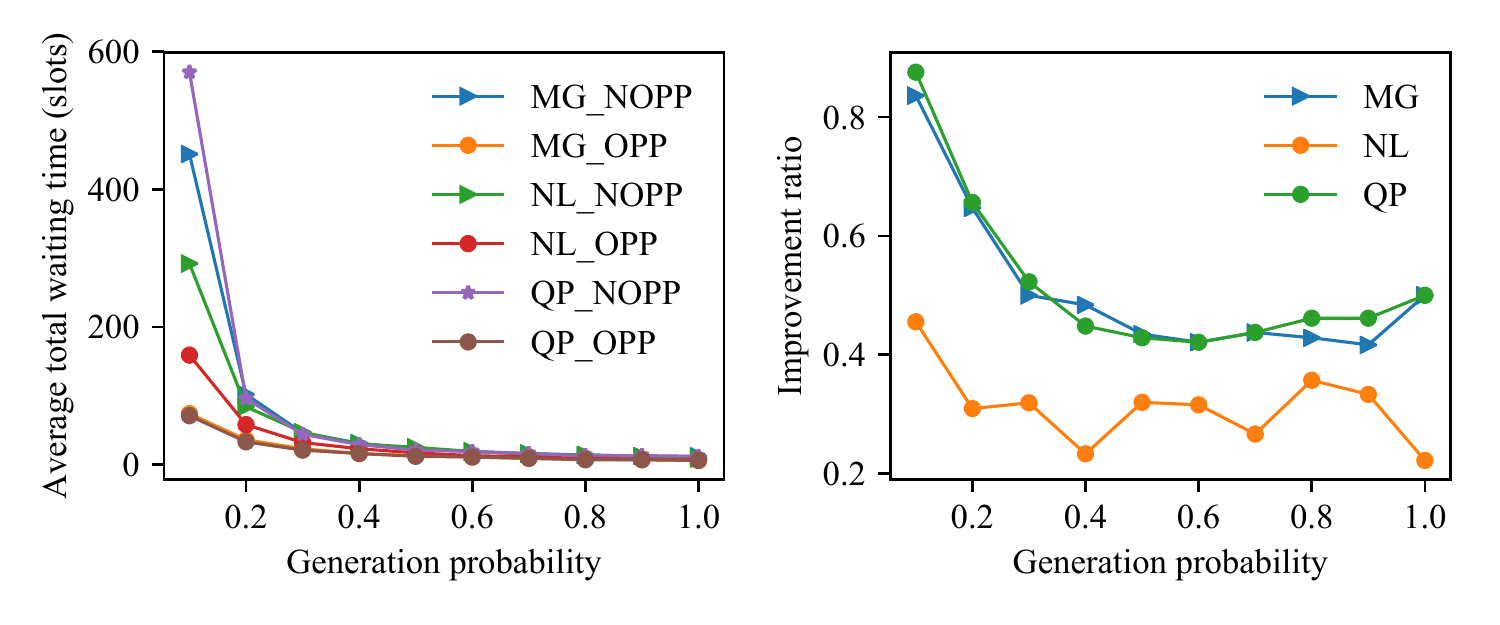}}
\caption{Average total waiting time in a grid network, with the fixed scenario $M=5$, $N=20$, $L=6$, $k=1$, and $p_\textsc{swap}=0.8$. The figure on the right shows the ratio of the improvement caused by opportunism to the non-opportunistic approach.}
\label{fig: results2}
\end{figure}

We now change the setup to another configuration, in which $M$, $N$ and $k$ stay the same, while we have $L=6$ and $p_\textsc{swap}=0.8$.
This new setup has much more dynamism than the previous one.
The results of comparing the average total waiting time of the three algorithms, based on $p_\textsc{gen}$, is shown in Fig.~\ref{fig: results2}.
We can see that in a high dynamism of the quantum network's state, the opportunistic approach enables significantly better performance, and this improvement is around 50\%.

\begin{figure}[b]
\centerline{\includegraphics[width=1\columnwidth]{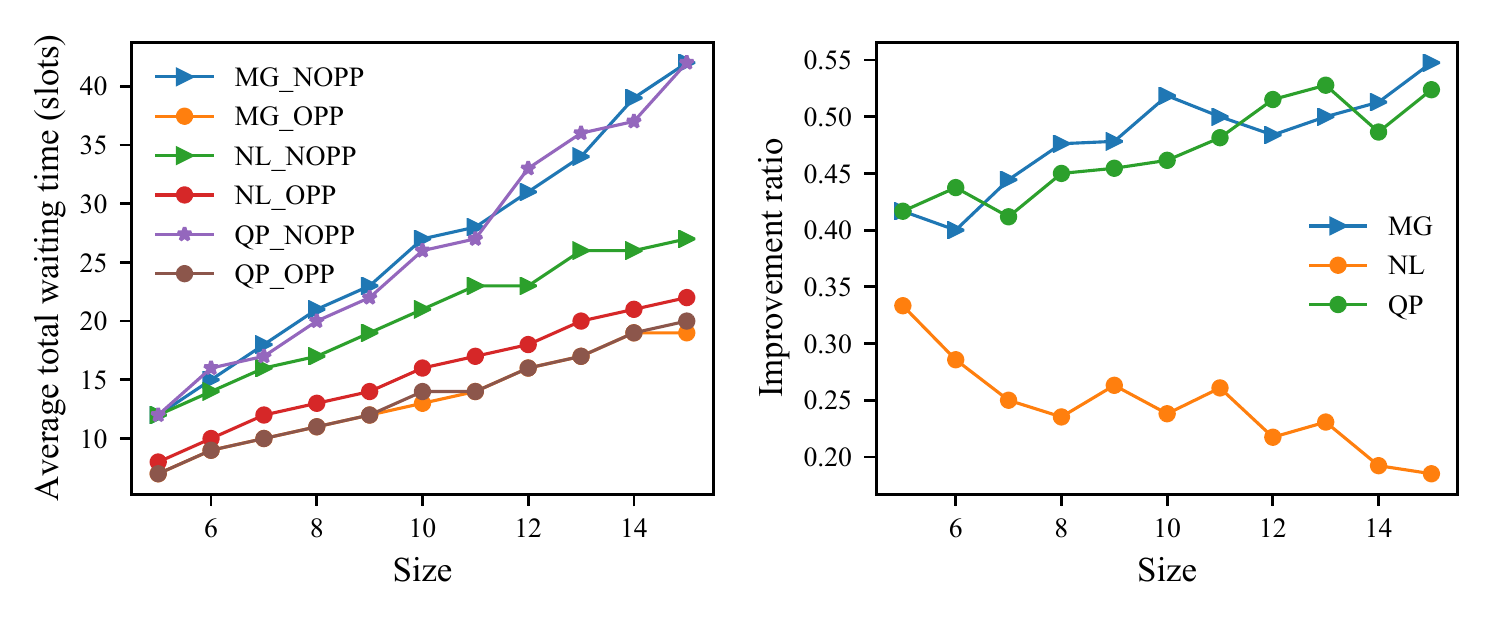}}
\caption{Average total waiting time in a grid network, with the fixed scenario $N=20$, $L=30$, $k=1$, $p_\textsc{swap}=0.8$, and $p_\textsc{gen}=0.8$. The figure on the right shows the ratio of the improvement caused by opportunism to the non-opportunistic approach.}
\label{fig: results3}
\end{figure}

Now we increase the network size and analyze the effect of opportunism in bigger networks.
Fig.~\ref{fig: results3} shows the average total waiting time based on the network size, for the three algorithms, in a grid network.
The setup is $N = 20$, $L = 30$, $k=1$, $p_\textsc{swap}=0.8$, and $p_\textsc{gen}=0.8$.
This setup is somehow an intermediate setup that is neither stable nor highly dynamic.
We can see that just like small networks, opportunism significantly improves performance in bigger networks as well.
\begin{figure}[htbp]
\centerline{\includegraphics[width=1\columnwidth]{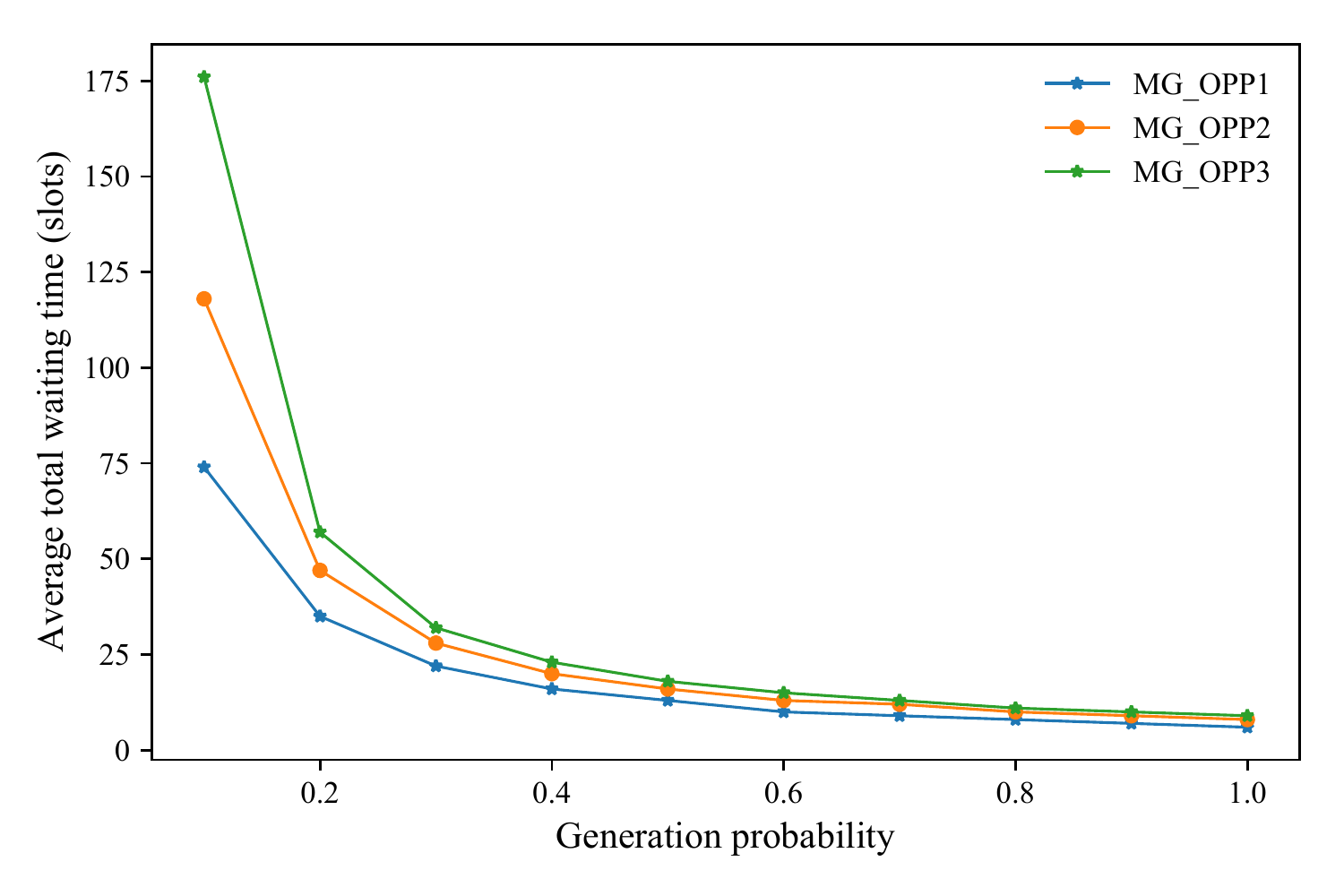}}
\caption{Average total waiting time in a grid network across different opportunism degrees, with the fixed scenario $N=20$, $L=30$, $p_\textsc{swap}=0.8$, and $p_\textsc{gen}=0.8$. The number after OPP is the opportunism degree.}
\label{fig: results4}
\end{figure}

We finally present Table~\ref{tab: ALWT}, in which we have shown the effect of opportunism on the average link waiting time.
The bottom row shows the improvement with respect to the non-opportunistic approach.
Due to lack of space, we have only included the MG algorithm.
Nonetheless, the improvement is similar in the other two algorithms.
The results clearly demonstrate the fact that opportunism decreases the average link waiting time, thus increasing the efficiency of resource consumption.
The setup in this part is $N = 20$, $M = 10$, $L = 30$, $p_\textsc{swap} = 1$, and $k=1$.
\begin{table}[!t]
\renewcommand{\arraystretch}{1.3}
\caption{Average link waiting time in the MG algorithm}
\label{tab: ALWT}
\centering
\begin{tabular}{c|c|c|c|c|c}
\hline
\bfseries $p_\textsc{gen}$ & 0.1 & 0.3 & 0.5 & 0.7 & 0.9\\
\hline\hline
NOPP & 12.9 & 8.8 & 7.18 & 6.3 & 5.3\\
\hline
OPP & 11.5 & 6.15 & 4.16 & 3.6 & 3.13\\
\hline
\bfseries Improv. & \bfseries 11\% & \bfseries 30\% & \bfseries 41\% & \bfseries 43\% & \bfseries 41\%\\
\hline
\end{tabular}
\end{table}
\par
\textbf{Degrees of Opportunism.}
In this part, we demonstrate the effect of changing the degree of opportunism.
The results verify our analysis, and as we can see in Fig.~\ref{fig: results4}, opportunism has a spectrum of choices for algorithm design.
Due to space limitations, we have only included the MG protocol here.
In Fig.~\ref{fig: results4}, the number after OPP is the opportunism degree.




\section{Related Work}\label{Sec: related work}
Opportunism, as studied in this paper, is related to several lines of work in quantum networks.
In terms of fundamental analysis, there are a series of works analyzing the limits of quantum communication, without examining the effect of opportunism. Below are several examples.
Pirandola et al.~\cite{FundamentalRepeaterless} analyze the fundamental limits of repeaterless communications using local operations and classical communication.
B\"{a}uml et al.~\cite{FundamentalBroadcast} study fundamental limitations on quantum broadcast networks by using multipartite entanglement instead of bipartite entanglement.
Patil et al.~\cite{DistanceIndependent} show that by leveraging n-qubit GHZ projective measurements, entanglement can be generated between two endpoints at a rate independent of the distance between them.
Khatri et al.~\cite{FigureMerit} and Bernardes et al.~\cite{RateAnalysisHybrid} analyze the average total waiting time for a group of links with infinite lifetimes, and Coopmans et al.~\cite{ImprovedBounds} provide analytical bounds on delivery times of long-distance entanglement in different protocols.
Praxmeyer~\cite{Reposition} analyzes the average total waiting time for a group of links with finite lifetimes.
\par
Routing in quantum networks has gained significant attention in the past few years. For example, 
Schoute et al.~\cite{ShortcutQuanRouting} analyze virtual links (i.e, entanglement between non-adjacent nodes) as primary resources for entanglement routing, in ring and sphere topologies.
Chakraborty et al.~\cite{DistRoutQuantumInternet} modify the greedy routing algorithm with regards to quantum networks, while also analyzing on-demand and continuous resource generation schemes.
Pant et al.~\cite{RoutingEntanglementInternet} introduce a multipath routing algorithm, utilizing local link information, in a grid topology.
Van Meter et al.~\cite{PathSelectionQuanRepeater} present several link costs and use Dijkstra's algorithm for calculating the cost of a path of links.
Shi et al.~\cite{ConcurrentRouting} propose using recovery paths in routing, present a nonlinear path cost function, and introduce an extended version of Dijkstra's algorithm for this function.
Dai et al.~\cite{OptimalProtocols} use linear programming to optimize the entanglement distribution rate between two endpoints, using repeaters with probabilistic swapping. However,
none of these algorithms consider opportunism as we did in this paper.
\par
\par
\section{Conclusion and Future Work}\label{sec: conclusion}
In this paper, we introduced opportunism in quantum networks and provided theoretical and experimental analysis with regard to its better performance when compared to the non-opportunistic approach.
We demonstrated the two inherent mechanisms by which opportunism increases efficiency, namely the swapping and reservation gains.
The results indicated that opportunism can decrease the average total waiting time up to 50\% in different setups and network dynamism.
Furthermore, we showed that opportunism is inherently flexible by introducing $k$-opportunism and demonstrating the spectrum of choices it provides.
Due to the importance of efficient resource allocation in quantum networks, the results of this paper demonstrate the need for a consistent approach towards integrating opportunism in the management layer of such networks (e.g., by including opportunism as a key functionality in the network layer of a quantum networking stack).
\par
Opportunism can be further analyzed in several directions.
The effect of opportunism on the fundamental entanglement distribution rate between two endpoints of a repeater chain needs further study.
Our analysis of the swapping and reservation gains does not include the finite lifetime and probabilistic swapping regime, and extending the results to include these assumptions would provide useful insight for designing new algorithms.
Furthermore, our analysis can be extended to more graph-theoretic results (including different topologies).
Finally, our simulator is currently supporting only grid networks and does not include the finite memory assumption.
Extending the simulator to cover these areas is also an interesting future work.


\section*{Acknowledgement}
\textcolor{black}{This work was supported by Natural Sciences and Engineering Research Council of Canada (NSERC) CREATE 528125-2019 and NSERC Discovery Grant RGPIN-2016-05310.}

\bibliographystyle{IEEEtran}
\bibliography{references}

\vspace{12pt}

\end{document}